\documentclass[aps,twocolumn,floatfix,amsmath,amssymb,prb]{revtex4-2}
\usepackage{graphicx}
\usepackage{color}

\usepackage{bm}
\usepackage{amsmath}

\DeclareMathOperator{\atanh}{atanh}
\DeclareMathOperator{\asinh}{asinh}

\begin{document}

\title{Noisy monitored quantum dynamics of ergodic multi-qubit systems}
\author{Henning Schomerus}
\affiliation{Department of Physics, Lancaster University, Lancaster, LA1 4YB, United Kingdom}

\begin{abstract}
I employ  random-matrix methods to set up and solve statistical models of noisy
nonunitary dynamics that appear in the context of monitored quantum systems.
The models cover a range of scenarios combining random dynamics and measurements of variable strength of one or several qubits.
The combined dynamics drive the system into states whose statistics reflect the competition of randomizing unitary evolution and the measurement-induced backaction collapsing the state.
These effects are mediated by entanglement,
as  I describe in detail by analytical results. For the paradigmatic case of monitoring via a single designated qubit,
this reveals a simple statistical mechanism, in which the monitoring conditions the state of the monitored qubit, which then imposes statistical constraints on the remaining quantities of the system.
For the case of monitoring several qubits with prescribed strength, the developed formalism allows one to set up the statistical description and solve it numerically. Finally, I also compare the analytical results to the monitored dynamics of a quantum kicked top, revealing two regimes where the statistical model either describes the full stationary dynamics, or resolves time scales during particular parts of the evolution.
\end{abstract}

\maketitle

The conceptual understanding of the role of measurements in quantum mechanics has significantly evolved over time \cite{vonNeumann1938MathematicalTheory, Wiseman2009QuantumControl,jacobs_2014}. Being rooted in a probabilistic description, a notable aspect of this discussion has been how the theory should be interpreted when applied to individual quantum systems, as well as individual parts of composed systems.
As for the understanding of measurements carried out on a subsystem, entanglement has been identified as a key signature that sets quantum systems apart from classical ones. In turn, the developments around experimental demonstrations of entanglement have shed significant light on the interpretation of the theory in individual realizations.
A central feature of entanglement is that  it is only useful if the state of the system is known to a sufficient extent. For instance, a random incoherent mixture of fully entangled states is completely indistinguishable from a mixture of separable states.
But with  advances in quantum optics and the recent advent of noisy intermediate scale quantum devices, quantum systems can be monitored and controlled in detail, justifying an approximate description by pure states even through the monitoring phase. Nonetheless, such monitoring leaves a significant impact on the dynamics, the backaction, whose early manifestation was the collapse of the wave function. This backaction differs from the quantum dynamics of the isolated system in that it is non-unitary, which results from the conditioning of the quantum state according to the measurement outcome.
The backaction is just as central to the description of measurements, as it means that monitoring is indeed useful---it entails that the measurement outcomes indeed reveal information about the quantum state, to the extent that in a complete measurement the outcomes pinpoint a definite, pure, post-measurement state. However, this requires to have access to all outcomes, as otherwise the measurement has the opposite effect of turning a pure state into an effectively mixed state.

These considerations enjoy a significant generalization when one considers the interaction of a quantum system with its environment---as pioneered, amongst others, by Fritz Haake, who developed comprehensive statistical descriptions of open quantum systems \cite{haake1973statistical}, and applied these both to specific paradigms as well as to advance the conceptual interpretation of the measurement process
(for a brief overview of this \oe{}uvre see \cite{Gnutzmann_2021}) . This laid the ground for our present understanding, in which many of the key aspects of measurements and open-system dynamics can be captured in a unified language. In this, entanglement with the environment gives rise to decoherence, which manifests itself in a reduced purity of the density matrix, meaning that the system effectively evolves into a mixed state. In turn, recording these interactions in sufficient detail can reveal enough information to protect the pure-state dynamics, as is manifested in weak, continuous, or variable strength measurements.

These conceptual advances also have guided the development of powerful theoretical approaches. As a statistical  tool, pure-state dynamics is useful as it can be averaged over incoherently to capture the general case of mixed-state dynamics, leading to frameworks such as quantum jump and trajectory formulations of unravelled master equations \cite{carmichael2009open}. And over the recent years, considering the pure-state dynamics in complex quantum circuits has brought about significant insights into the interplay of dynamical entanglement generation and spreading and the reduction of entanglement in local measurements \cite{Chan2019Unitary-projectiveDynamics, Skinner2019Measurement-InducedEntanglement,  Li2018QuantumTransition, Li2019Measurement-drivenCircuits, Li2021, Gullans2019purification, Gullans2020, Zabalo2020CriticalCircuits, PhysRevB.104.155111, Zabalo2022, Li2020Conformal, Bao2020, Jian2020, Fan2021, Bao2021, Bera2020, Sang2021measurement, Zhang2020, Choi2020, Nahum2021, Rossini2020, Iaconis2020,Kalsi2022Threefold}. This revealed a measurement-induced phase transition in the entanglement entropy, which changes from an extensive to an intensive quantity (a volume law to an area law) when the measurement strength exceeds a certain value. While  initially encountered in a stroboscopic setting with maximally random local dynamics and hard measurements, this transition is now known to occur in a wide range of related settings, and in particular, also for continuously evolving systems monitored with variable strength \cite{Szyniszewski2019EntanglementMeasurements,Szyniszewski2020UniversalityEntanglement}.

It is against the backdrop of these systems that I formulate, in this paper, a statistical description of noisy monitored pure-state dynamics in multi-qubit systems. In contrast to the works concerning the entanglement transition, I here consider noisy dynamics in which all qubits interact on equal terms.
Furthermore, instead of monitoring all of the qubits, I focus on the case that this is done for a specific designated subset of the qubits, which are subjected to variable-strength measurements. In absence of the monitoring, the dynamics displays universal behaviour in accordance to random-matrix theory, another field of significant activity in Fritzens  work \cite{Haake2018QSoC}. This sets a benchmark to investigate how monitoring the qubits affects the quantum state of the complete system, revealing the role of correlations and entanglement in mediating the backaction to other parts of the system. I also take the opportunity to apply these results to a specific physical setting, the monitored dynamics of a kicked quantum top \cite{haake1987classical}, the paradigmatic model that Fritz introduced to study how complex systems display the quantum signatures of chaos \cite{Haake2018QSoC}.

The underlying physics is most clearly revealed in noisy systems that are monitored with variable strength via a single designated qubit. In this case, we can obtain comprehensive analytical results for the statistical features of the quantum state in the stationary limit of long times.
In this statistical description,  it turns out that all nontrivial effects of the monitoring on the dynamics can then be captured by a single quantity, and the geometric constraints that this quantity poses on others.
This quantity, which we will denote as $r$, simply represents the probability that the monitored qubit is in the state $|0\rangle$ in the monitoring basis.
The underlying physical picture is that besides the monitoring-induced conditioning on $r$, the quantum state of the composed system is statistically free. The intriguing feature is how powerful these $r$-dependent constraints are in transferring the effects of the monitoring to other parts of the system.
This results in highly complex and characteristic statistical signatures of the monitoring as a function of its strength, which we reveal by adopting parameterizations that highlight the role of correlations and entanglement.
These features also extend to the case of monitoring several of the qubits with possibly distinct strength, where results can be obtained by numerical solution of the statistical model.
Finally, the comparison of the analytical results of the monitored noisy dynamics to the monitored dynamics of the quantum kicked top \cite{haake1987classical} reveals  two regimes,
where the statistical model
either describes the full stationary dynamics in the kicked top, or resolves the deterministic dynamical time scales during particular parts of its evolution.

This work is organised as follows. Section \ref{sec:setting} describes the general setting and the properties of the quantum state we are interested in, and sets out the monitoring protocol.
 Section \ref{sec:freedynamics} addresses the noisy dynamics in the absence of monitoring, which serves as a benchmark for the monitored dynamics, and allows us to develop the general methodology, which is closely related to Dyson's Brownian motion approach \cite{Dyson1962Brownian}.
With Section \ref{sec:onemonitoredqubit}, we turn to the instructive case of multi-qubit systems monitored by a single designated qubit, for which we can obtain exact analytical results.
Section \ref{sec:multiple} complements this with numerical results for systems with multiple monitored qubits.
In Sec.~\ref{sec:kickedtop}, I then compare the analytical results of Sec.~\ref{sec:onemonitoredqubit}~ to the monitored dynamics of a kicked top.
The results are summarised in the concluding Sec.~\ref{sec:conclusions}, which also describes further implications and gives an outlook on possible extensions.

\section{Setting}
\label{sec:setting}
We are interested in the dynamics of quantum systems under repeated steps of random unitary evolution and variable-strength measurements.
This section describes the types of systems along with  their quantities of interest, as well as the applied measurement protocol. In the next sections this is then applied to various dynamical scenarios.

\subsection{Systems and quantities of interest}
We consider the pure-state dynamics $|\psi(t)\rangle$ of quantum systems with a finite Hilbert-space dimension $N$,
which will come about by the combination of unitary time evolution and measurement steps with recorded outcomes.
We denote the amplitudes in the natural computational basis $|n\rangle$ as $\psi_n=\langle n | \psi \rangle$,
and will be particularly interested in the statistics of the normalised probability coefficients
\begin{equation}
r_n=|\psi_n|^2, \quad \sum_n r_n=1
\label{eq:rn}
\end{equation}
attained at large times.

Our main focus will be on systems made out of one or several two-level systems (here referred to as qubits), where the Hilbert space is a tensor product of $q$ two-dimensional spaces, and overall $N=2^q$, where we let indices run from $n=1$ to $N$.
For a single-qubit system, we write explicitly
\begin{equation}
|\psi\rangle=\alpha|0\rangle+\beta|1\rangle
\end{equation}
and set
\begin{equation}
r=|\alpha|^2=1-|\beta|^2.
\label{eq:r}
\end{equation}

For a two-qubit system, we write
\begin{equation}
|\psi\rangle=\alpha|00\rangle+\beta|01\rangle+\gamma|10\rangle+\delta|11\rangle
\end{equation}
and set
\begin{align}
r&=|\alpha|^2+|\beta|^2=1-|\gamma|^2-|\delta|^2,\\
R&=|\alpha|^2+|\gamma|^2=1-|\beta|^2-|\delta|^2,\\
C&=|\alpha|^2|\delta|^2-|\beta|^2|\gamma|^2.
\label{eq:rRc0}
\end{align}
Here $r$ and $R$ are the probabilities to find the first or second qubit in the state $|0\rangle$. The quantity $C$ quantifies the experimentally observable correlations of the measurement outcomes specifically carried out in the computational basis, and will feature naturally in the statistical analysis. For this, we will make use of the following relations,
\begin{align}
&|\alpha|^2=C+rR, \quad |\beta|^2=r(1-R)-C,
\\ & |\gamma|^2=R(1-r)-C, \quad |\delta|^2=(1-r)(1-R)+C.
\label{eq:rRc}
\end{align}
Furthermore, we will also consider the concurrence
\begin{equation}
\mathcal{C}=2|\alpha\delta-\beta\gamma|,
\end{equation}
which quantifies the degree of entanglement as detectable by combining measurements inside and outside the computational basis.

For a system of $q$ qubits, we work with the probability coefficients $r_n$ defined in Eq.~\eqref{eq:rn}, as well as with the
probability coefficients
\begin{equation}
R_m=\sum_{n|\text{qubit $m$ in state $|0\rangle$}}r_n
\label{eq:Rgendef}
\end{equation}
of the individual qubits, and where convenient abbreviate
\begin{align}
r&\equiv R_1=\sum_{n=1}^{N/2}r_n,
\nonumber
\\
R&\equiv R_2=\sum_{n=1}^{N/2}r_{2n-1},
\label{eq:rshortcuts}
\end{align}
for the probability to find the first  or second qubit in state $|0\rangle$.

\subsection{Measurement protocol}

We implement  measurements of variable strength $\lambda$ by controlled  entanglement with a suitable auxiliary system, whose state is then measured and the outcome recorded.
The measurements are assumed to occur instantaneously, where pre-measurement states are denoted by
a time index $t_-$ and post-measurement states by a time index  $t_+$.
Depending on the measurement protocol, this can be designed to either extract information discriminating all states of the system, or only those of a subsystem.
We here set this out for the case of one and two qubits, and refer to Appendix \ref{app:b} for a general version of this protocol for a system of $N$ states.

\subsubsection{Description for a single qubit}

Denoting the normalised state of a single qubit as
\begin{equation}
|\psi(t_-)\rangle=\alpha|0\rangle+\beta|1\rangle,
\end{equation}
we first entangle it with an ancilla qubit. This is designed to take the initially separable joint state
\begin{equation}
|\varphi\rangle=|\psi(t_-)\rangle\otimes|0\rangle
\end{equation}
into the intermediate state
\begin{align}
&|\varphi\rangle=\alpha |0\rangle\otimes|a_\lambda\rangle+\beta |1\rangle\otimes|a_{-\lambda}\rangle,
\\
& |a_{\lambda}\rangle=\sqrt{\frac{1+\lambda}{2}}|0\rangle+\sqrt{\frac{1-\lambda}{2}}|1\rangle,
\end{align}
where $\lambda\in[0,1]$ determines the measurement strength.
In a quantum circuit, this step can be implemented by a conditional gate operation.
Measuring the ancilla in its computational basis brings the system back into a separable state, which depends on the measurement outcome. We use an index
$\eta=\pm1$ to distinguish the  two possibilities, $\eta=1$ if the ancilla is found in state $|0\rangle$ and $\eta=-1$ if the ancilla is found in state $|1\rangle$. Post-measurement, the qubit of interest then assumes the state
\begin{equation}
|\psi_\eta(t_+)\rangle=\frac{\sqrt{1+\eta\lambda}\alpha|0\rangle+\sqrt{1-\eta\lambda}\beta|1\rangle}{\sqrt{1+\eta\lambda(|\alpha|^2-|\beta|^2)}}
.
\label{eq:postm}
\end{equation}
These outcomes occur with probabilities
\begin{equation}
P_\eta=\frac{1+\lambda\eta(|\alpha|^2-|\beta|^2)}{2}.
\label{eq:peta}
\end{equation}

The effect that the quantum state changes upon the measurement is generally know as backaction.
According to \eqref{eq:postm}, this here modifies the probability coefficient to
\begin{equation}
r_{t_+}=\frac{(1+\eta\lambda) r_{t_-}}{1+\eta\lambda(2 r_{t_-}-1)}.
\end{equation}
Averaged over the measurement outcomes $\eta=\pm1$, we then always have
\begin{equation}
\overline{r_{t_+}}=r_{t_-}.
\end{equation}

For measurement strength $\lambda=1$, this description reduces to the standard von Neumann protocol of projective measurements, while general values $0<\lambda<1$ result in variable-strength measurements.
If $\lambda$ is small, we can approximate the backaction by the following simplified description,
\begin{align}
&r_{t_+}= r_{t_-}+dr_{t_-},
\\
&dr\equiv 2\xi\lambda r(1-r),
\label{eq:dr}
\end{align}
where $\xi=\pm1$ again signals random measurement outcomes, but these are now taken to occur with equal, state-independent probability $P(\xi=\pm1)=1/2$.
As indicated in our notation, we defined the increments so that they refer to the initial data (Ito calculus), and in a way where we can then drop the time index.
The increment $dr$ correctly reproduces the first and second moment of the backaction on the coefficient in the leading orders of the measurement strength,
\begin{align}
&\overline{dr}=0+O(\lambda^3),
\\
&\overline{(dr)^2}=4\lambda^2r^2(1-r^2)+O(\lambda^3).
\end{align}
When interpreted stochastically, these equations can be used to describe the continuous monitoring of the system, which is the regime that we will focus on in this paper. In this case, the  measurement backaction effectively scales as $\lambda^2$, which we later quantify relative to a similar scaling of noisy unitary dynamics in the considered systems.

\subsubsection{Description for two qubits}
\label{sec:twoqubitmonitoringprotocol}
The protocol for a single qubit can be extended to multiple qubits in several ways.
Here, we describe the situation that we monitor one of the qubits. Because observables of distinct qubits commute, the case where both qubits are monitored then simply follows from composition of these measurements.
The closely related  case to monitor the composed system with an auxiliary 4-level system is covered by the general protocol for $N$-level systems, described in Appendix \ref{app:b}.

We designate the first qubit as the monitored one, which is again achieved by utilizing an ancilla. The physical entanglement and measurements steps are the same as for the isolated qubit,
resulting in the post-measurement state
\begin{equation}
|\psi_\eta(t_+)\rangle=\frac{\sqrt{1+\eta\lambda}}{\sqrt{2P_\eta}}(\alpha|00\rangle+\beta|01\rangle)+
\frac{\sqrt{1-\eta\lambda}}{\sqrt{2P_{-\eta}}}(\gamma|10\rangle+\delta|11\rangle),
\end{equation}
where the outcomes $\eta=\pm1$ now occur with probability
\begin{equation}
P_\eta=\frac{1+\lambda\eta(|\alpha|^2+|\beta|^2-|\gamma|^2-|\delta|^2)}{2}.
\label{eq:peta2}
\end{equation}

In terms of the parameterization \eqref{eq:rRc},
this gives
\begin{align}
r_{t_+}&=\frac{(1+\eta\lambda) r_{t_-}}{1+\eta\lambda(2 r_{t_-}-1)},
\\
R_{t_+}&=R_{t_-}+\frac{2C_{t_-}\eta\lambda}{1+\eta\lambda(2 r_{t_-}-1)},
\\
C_{t_+}&=C_{t_-}
\frac{1-\lambda^2}{(1+\eta\lambda(2 r_{t_-}-1))^2}.
\end{align}
We see that the backaction also affects the probability coefficient $R$ of the other qubit, and that this is mediated via the correlation parameter $C$.

For small measurement strength $\lambda$, we can again obtain the backaction from a simplified stochastic process, with increments
\begin{align}
dr&=2\xi\lambda r(1-r),
\\
dR&=2\xi\lambda C,
\\
dC&=2\lambda C((1-2r)\xi-2\lambda r(1-r))
\end{align}
and $\xi=\pm1$ again occurring with equal, state-independent probability $P(\xi=\pm1)=1/2$.

An analogous process is obtained when we monitor the second qubit with strength $\lambda'$ and monitoring noise $\xi'$, and both can be combined into a single process
given by
\begin{align}
dr&=2\xi\lambda r(1-r)+2\xi'\lambda' C,
\\
dR&=2\xi'\lambda' R(1-R)+2\xi\lambda C,
\\
dC&=2C[\xi\lambda (1-2r)  -2\lambda^2 r(1-r)
\nonumber\\
&\quad
+ \xi'\lambda'(1-2R)-2{\lambda'}^2 R(1-R)]
.
\end{align}
This composition is valid because the noise variables $\xi$ and $\xi'$ are independent of each other, and analogous considerations will allow us to extend these descriptions further to include noisy unitary dynamics.

\subsection{The role of monitoring}

If we do not have access to the  outcomes, the measurements induce decoherence and drive the system into a mixed state, with the decoherence rate governed by $\lambda$ (in the continuous case, scaling as $\lambda^2$).
In the context of monitored dynamics, however, it is assumed that the outcome is recorded, so that the system can always be described as a pure state.
In this case we follow the quantum state through random sequence of outcomes, occurring with the  probabilities such as those stated in Eqs. \eqref{eq:peta} and \eqref{eq:peta2}. The measurements introduce both nonlinearity and randomness, with the former occurring due to the normalization in the probabilities, while the latter supplements the randomness from the unitary dynamics to which we turn now.

\section{Noisy unitary dynamics of the isolated system}
\label{sec:freedynamics}
We let the noisy dynamics unfold in eventually infinitesimally small time steps $dt$, indexed by a discrete time $\tau$ on the quantum state $|\psi_\tau\rangle\equiv|\psi(\tau dt)\rangle$.
In the absence of measurements, the isolated system dynamics can therefore be written as
\begin{equation}
|\psi_{\tau+1}\rangle=U_\tau|\psi_{\tau}\rangle,
\label{eq:discdyn}
\end{equation}
and hence is obtained from the time-evolution operator $U_\tau$ over the given time step,
which is an $N\times N$ dimensional unitary matrix.
Aiming at a statistical description of a noisy system,
we introduce randomness into these dynamics by a standard random-matrix approach, mirroring Dyson's Brownian motion  \cite{Dyson1962Brownian} and taking the general form of a Wiener process.  In this section, we define this process in detail for the isolated system, and obtain the resulting statistics of the quantum state at long times. This recovers results connected to the circular unitary ensemble (CUE) of random-matrix theory, but adapted to the quantities of interest, and will serve as a benchmark for the monitored dynamics. Furthermore, discussing the isolated dynamics first allows us the introduce general methodology that we then can extend to the monitored case.

\subsection{Wiener process}
We generate the random noisy dynamics from
time evolution operators that are close to the identity,
 \begin{equation}
U_\tau=1+i \varepsilon H_\tau-\varepsilon^2/2+O(\varepsilon^{3}),
\label{eq:wienergen}
\end{equation}
where the dimensionless generators $H_\tau$ are taken independently from the Gaussian unitary ensemble (GUE), scaled so that in the ensemble average $\overline {H_\tau^2}=\openone$.  The generators then play the role of noise in a Wiener process,
as quantified by the ensemble-averaged strength
\begin{equation}
\overline{|(U_{\tau})_{nm}|^2}= \varepsilon^2/N
\end{equation}
 of the off-diagonal elements $n\neq m$. Consequently, the dynamics unfold on an effective time scale $t_\varepsilon= dt/\varepsilon^2$.
Interpreted parametrically as a process sampling $U(n)$, hence considering  the composed time-evolution operator $U=\prod_\tau U_\tau$ itself, this amounts to  Dyson's Brownian motion process, a powerful tool to obtain statistical insights into the CUE whose Haar measure is approached in the stationary limit.
This  process is then conveniently studied using  Fokker-Planck equations, as we set out here in terms of the quantum-state dynamics itself.

For a given set of statistical quantities $z_n=x, y, \ldots$, the passage to the Fokker-Planck equation requires us to obtain the drift, diffusion, and cross-correlation coefficients
\begin{align}
\label{eq:driftgen}
&d_x\equiv \lim_{\varepsilon\to 0}\frac{\overline{x_{\tau+1}-x_\tau}}{dt/t_\varepsilon}\\
\label{eq:diffgen}
&D_x\equiv  \lim_{\varepsilon\to 0}\frac{\overline{(x_{\tau+1}-x_\tau)^2}}{dt/t_\varepsilon}\\
\label{eq:crossgen}
&D_{x,y}\equiv  \lim_{\varepsilon\to 0}\frac{\overline{(x_{\tau+1}-x_\tau)(y_{\tau+1}-y_\tau)}}{dt/t_\varepsilon},
\end{align}
obtained from the ensemble-averaged first and second moments of the increments in order $\varepsilon^2$.
In analogy to our convention for the measurements, we express these coefficients again in terms of the initial data $x_\tau, y_\tau, \ldots$ (Ito calculus), and in the interest of compact notation then drop the time index $\tau$.
The corresponding Fokker-Planck equation takes the general form
\begin{align}
\frac{\partial}{\partial t} P(\{z_k\};t)&=-\sum_n \frac{\partial}{\partial z_n}d_{z_n}  P(\{z_k\};t)
\nonumber\\
&+\frac{1}{2}
\sum_{nm}\frac{\partial^2}{\partial z_n\partial z_m}D_{z_n,z_m} P(\{z_k\};t),
\end{align}
where we identify $D_{z_n,z_n}=D_{z_n}$.
The stationarity statistics $P(\{z_k\})=\lim_{t\to\infty}P(\{z_k\};t)$ are then obtained from the stationarity condition $\frac{\partial}{\partial t} P(\{z_n\};t)=0$.
We now apply this approach to systems with different numbers of qubits.

\subsection{Noisy dynamics of a single qubit}
For a system with a single qubit parameterized as in Eq.~\eqref{eq:r}, we find the ensemble averages
\begin{align}
&d_r=1/2-r,\\
&D_r= r(1-r).
\end{align}
These give rise to the Fokker-Planck equation
\begin{equation}
\frac{\partial}{\partial t} P(r;t)=-t_\varepsilon^{-1}\frac{\partial}{\partial r}\left[\left(\frac{1}{2} - r- \frac{1}{2}\frac{\partial}{\partial r}r(1-r)\right)  P(r;t)\right]
\end{equation}
for the time-dependent probability density $P(r;t)$. The transient dynamics depends on the initial distribution, which for instance may represent a random or a prescribed state.  Here we focus on the emergent stationary behaviour at large times, which follows from the stationarity condition $\frac{\partial}{\partial t} P(r;t)=0$.
In the present setting, the resulting stationary distribution is uniform,
\begin{equation}
P(r)=1\quad (0\leq r\leq1).
\label{eq:pr0}
\end{equation}
Monitoring the system will modify this distribution due to the measuring backaction.

\subsection{Noisy dynamics of multiple qubits}

For  systems of $q$ qubits, with states parameterized as in Eq.~\eqref{eq:r}, we similarly find
\begin{align}
&d_{r_n}=1/N-r_n,\\
&D_{r_n}=\frac{2}{N} r_{n}(1-r_{n}),\\
&D_{r_n,r_m}=-\frac{2}{N} r_{n}r_{m},
\end{align}
with the cross-correlator in the last line applying to $n\neq m$.
The expressions for each $r_n$ close, so that their marginal distributions follow directly from the corresponding Fokker-Planck equation
\begin{align}
&\frac{\partial}{\partial t} P(r_n,t) =
\nonumber\\
&\quad
-t_\varepsilon^{-1}\frac{\partial}{\partial r_n}\left[\left(\frac{1}{N}-r_n + \frac{1}{N}\frac{\partial}{\partial r_n}r_n(1-r_n)\right)  P(r_n,t)\right].
\end{align}
From the stationarity condition, these distribution are then found to be given by
\begin{equation}
P(r_n)=(N - 1) (1 - r_n)^{N-2} \quad (0\leq r_n\leq1).
\end{equation}
These distributions are again independent of the initial conditions, and correspond to the known distributions of individual matrix elements in the CUE, which reflects the unitary invariance of this ensemble. This invariance will no longer hold true in the presence of monitoring, so that these distributions serve as a useful benchmark.

\subsection{Two qubits and the role of constraints}
For two qubits,  the four probability coeffients $r_n$, $n=1,2,3,4$ follow the parabolic distribution
\begin{equation}
P(r_n)=3 (1 - r_n)^2 \quad (0\leq r_n\leq1).
\label{eq:rntwoq}
\end{equation}

The parameterization \eqref{eq:rRc}  for the state of two qubits leads to statistical quantities that are not naturally discussed in the context of CUE matrix statistics, even though they can be obtained from it (see Appendix \ref{app:rRc}).
For these quantities we find
\begin{align}
&d_r=1/2-r,
\nonumber\\
&d_R=1/2-R,
\nonumber\\
&d_C=-3C/2+ (1/2 - r)(1/2 - R),
\nonumber\\
&D_r=r(1 - r)/2,
\nonumber\\
&D_R=R(1 - R) /2,
\nonumber\\
&D_C=[C (1 - 2 r) (1 - 2 R)-C^2+r(1 - r) R (1 - R)]/2 ,
\nonumber\\
&D_{r,R}=C/2,
\nonumber\\
&D_{r,C}=(1/2 - r)C,
\nonumber\\
&D_{R,C}=(1/2 - R)C.
\label{eq:rrcddfree}
\end{align}
These coefficients can again be introduced into a corresponding Fokker-Planck equation.
Remarkably, as
\begin{equation}
\sum_{x\in\{r,R,C\}} \frac{\partial}{\partial x} d_x=\frac{1}{2}\sum_{x,y\in\{r,R,C\}} \frac{\partial^2}{\partial x\partial y} D_{x,y},
\end{equation}
 the resulting stationary joint distribution of these quantities is still uniform, but is subject to nontrivial constraints on the domain of $C$:
\begin{align}
P(r,R,C)=6, \quad \Big( & 0\leq r,R \leq 1 \mbox{ and}
\nonumber
\\
&  -\min((1-r)(1-R),r R)<C
\nonumber
\\
& \quad<\min(r(1-R) ,R(1-r))\Big).
\label{eq:prRc}
\end{align}

\begin{figure}
  \centering
\includegraphics[width=0.8\columnwidth]{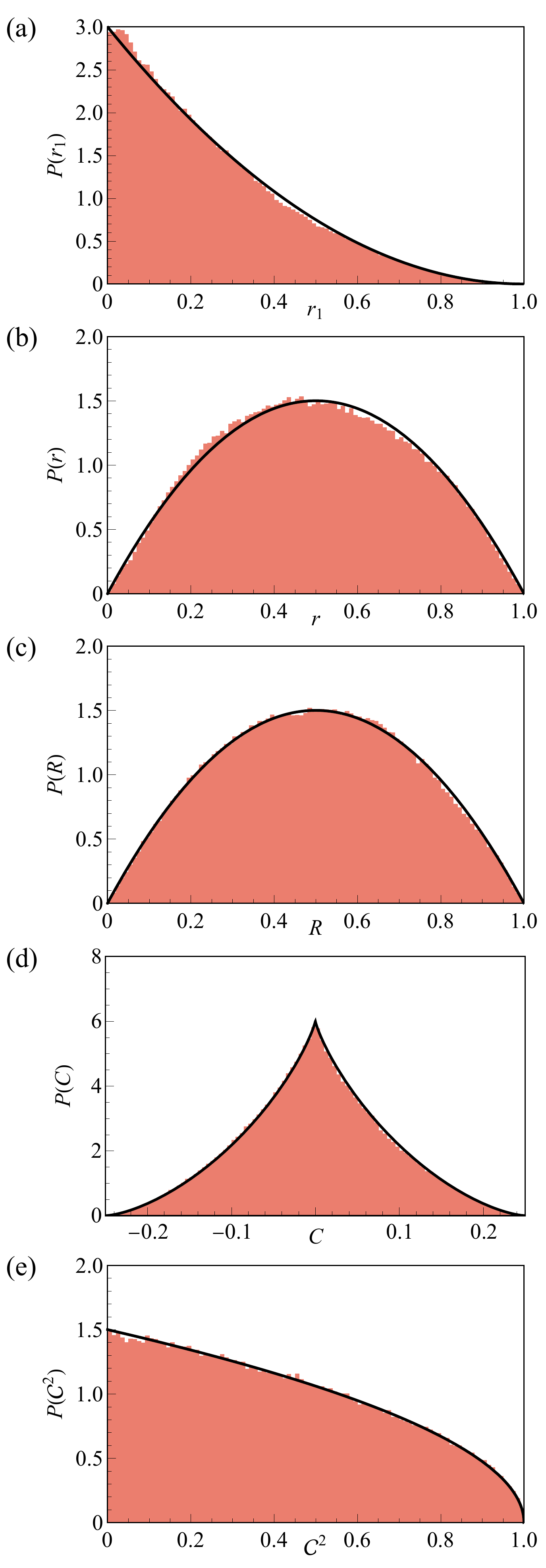}
  \caption{For an isolated system of two qubits evolving under noisy unitary dynamics
\eqref{eq:discdyn} generated by unitary operators \eqref{eq:wienergen}, the panels compare analytical results (curves) for the stationary statistics of the quantum state to numerical data (histograms) from a single trajectory obtained for $\varepsilon=0.1$.
The statistical quantities and analytical results are (a) probability coefficient $r_1$ to find the qubits in state $|00\rangle$, Eq. \eqref{eq:rntwoq}, (b,c)  probability coefficients $r$ and $R$ to find the first or second qubit in state $|0\rangle$, Eqs. (\ref{eq:prtwoq},\ref{eq:pRtwoq}), (d) correlation coefficient $C$,   Eq.~\eqref{eq:pctwoq}, and (e) squared concurrence $\mathcal{C}^2$, Eq.~\eqref{eq:pconctwoq}.}
\label{fig:oneoftwofree}
\end{figure}

These constraints have a geometric interpretation, where they relate to extremal positions of state vectors of lengths $r$ and $R$ referring to parts of the quantum state where the first or second qubit is in state $|0\rangle$.
When one integrates out the complementary variables, the constraints determine the integration domain, which leads to the marginal distributions
 \begin{align}
 P(r)&=6r(1-r) \quad (0<r<1),
\label{eq:prtwoq}
 \\
 P(R)&=6R(1-R) \quad (0<R<1),
\label{eq:pRtwoq}
 \\
 P(C)&=6\sqrt{1-4|C|} - 24 |C| \atanh(\sqrt{1-4|C|})
 \nonumber
 \\ & \qquad (-1/4<C<1/4).
\label{eq:pctwoq}
 \end{align}
As the drift and diffusion expressions for $r$ and $R$ again close individually, their marginal  contributions can also be obtained directly from the corresponding Fokker-Planck equations, delivering the same results.
We can also write down the joint distribution of the two probability coefficients,
\begin{equation}
P(r,R)=6\min(r,R,1-r,1-R), \quad (0\leq r,R \leq 1).
\end{equation}

Furthermore, we find that the equations for the concurrence $\mathcal{C}$ also close,
\begin{align}
&d_{\mathcal{C}}= \frac{1}{4 \mathcal{C}}-\mathcal{C},
\\
&D_{\mathcal{C}}=\frac{1}{2}(1-\mathcal{C}^2).
\end{align}
This results in the distribution
\begin{equation}
P(\mathcal{C})=3\mathcal{C}\sqrt{1-\mathcal{C}^2},
\label{eq:pconctwoq}
\end{equation}
or, equivalently
\begin{equation}
P(\mathcal{C}^2)=\frac{3}{2}\sqrt{1-\mathcal{C}^2}.
\label{eq:pconctwoqa}
\end{equation}

In Figure \ref{fig:oneoftwofree}, these analytical results are compared to numerical results. While for the present case of isolated unitary dynamics such data could be obtained by directly sampling the CUE, this would not extend to the case of monitored dynamics. Therefore, we base the numerics on the discretised stochastic time evolution \eqref{eq:discdyn}, generated by unitary operators \eqref{eq:wienergen} with $\varepsilon=0.1$.
The data in Figure \ref{fig:oneoftwofree} is then obtained from the states along a single trajectory, of length $5\times 10^5$ time steps. This results in good agreement, which we next aim to replicate for monitored dynamics.
This will also guide us to a more general understanding of geometric constraints such encountered in Eq.~\eqref{eq:prRc}.

\section{Dynamics monitored by a single designated qubit}
\label{sec:onemonitoredqubit}

We now combine the noisy unitary dynamics with continuous monitoring, following a protocol where each unitary step of the time evolution is supplemented by a variable-strength measurement, both considered over an infinitesimally small time step.  As the noise in these two stochastic processes is mutually uncorrelated, they can be combined into a single process in which drift, diffusion, and cross-correlation coefficients sum up. Maintaining the definitions \eqref{eq:driftgen}, \eqref{eq:diffgen}, and \eqref{eq:crossgen} of these coefficients including their scaling with $t_\varepsilon$, the dynamics are then governed by a dimensionless monitoring strength $\Lambda=\lambda^2/\varepsilon^2$.
This allows us again to formulate Fokker Planck equations for the quantities of interest, which we can study in the stationary limit. In the present section, we carry  out this program for the case of a system in which one designated qubit is monitored.

\subsection{Single-qubit system}

We start with the simplest case, in which the monitored qubit is the only qubit in the system.
Combining the noisy unitary dynamics with continuous monitoring of this qubit, the probability coefficient $r$ obtains the drift and diffusion coefficients
\begin{align}
&d_r=1/2-r,\\
&D_r= r(1-r)+4\Lambda r^2(1-r^2).
\end{align}
This gives the joint distribution
\begin{equation}
P(r)= \frac{c_\Lambda}{\left(1+4 \Lambda r(1-r)\right)^2},
\end{equation}
where $c_\Lambda$ is for normalization.
We see that the monitoring induces a bimodal character to this distribution, which becomes concentrated at $r=0$ and $r=1$ as the effective monitoring strength $\Lambda$ increases. Physically, this can be interpreted as a signature of the measurement backaction.

\subsection{Monitoring of one of two qubits}
\label{sec:mononeoftwo}

\begin{figure}
  \centering
  \includegraphics[width=\columnwidth]{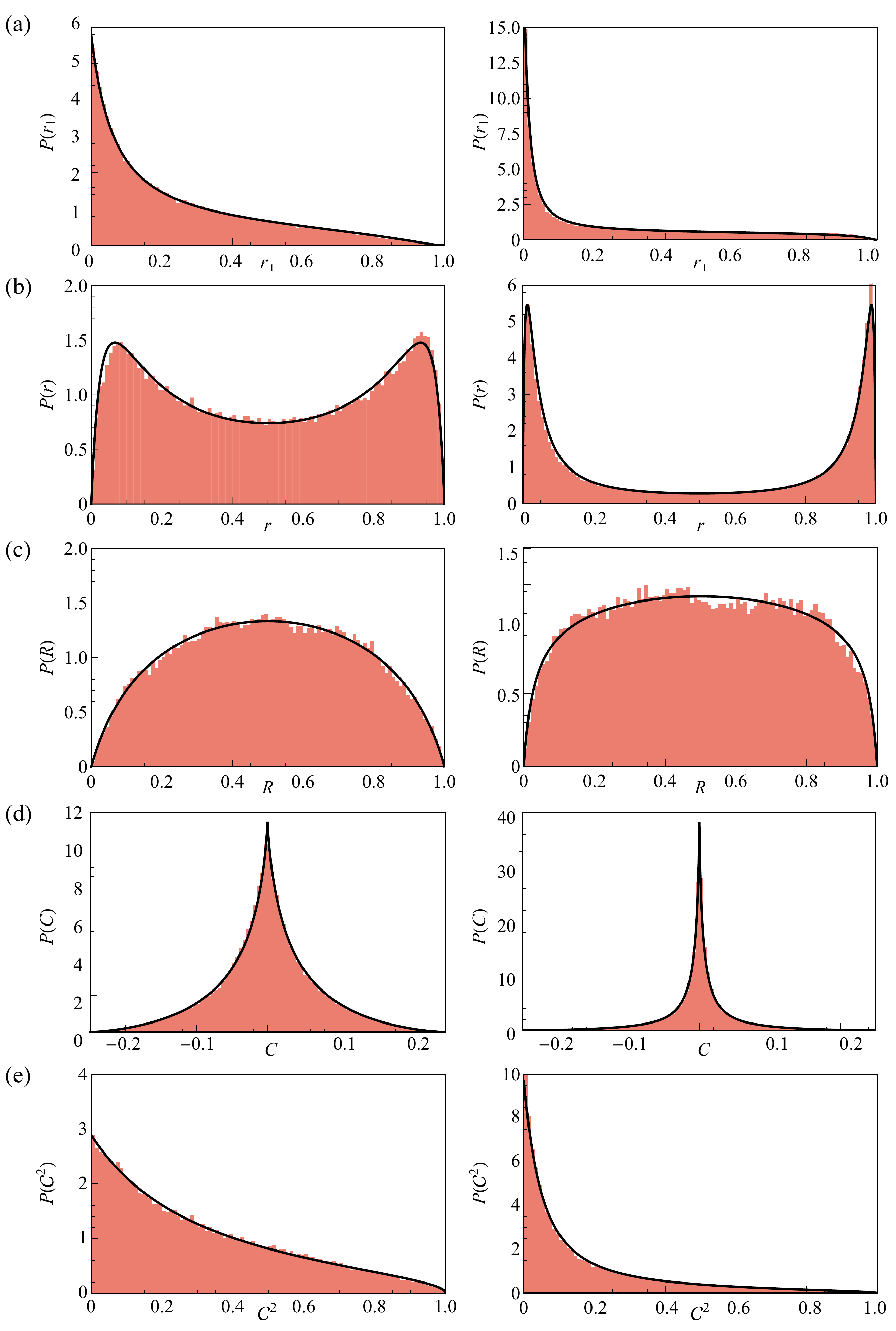}
  \caption{Stationary quantum state statistics in a system of two qubits, where the first qubit is monitored with effective strength $\Lambda=1$ (left column) or  $\Lambda=5$ (right column). The different panels address the same quantities as in Fig.~\ref{fig:oneoftwofree}, and the numerical data is obtained in an analogous way, but with measurements of strength $\lambda=0.1$ (left column) or  $\lambda=\sqrt{0.05}$ (right column) carried out after each unitary step. The curves represent the analytical expressions from Sec.~\ref{sec:mononeoftwo}
 and Appendix \ref{app:longexpressions}.}\label{fig:oneoftwo}
\end{figure}

We next apply these considerations to a system of two qubits, where the first one is designated to be monitored.
This situation is usefully studied in the parameterization \eqref{eq:rRc}, where we obtain
\begin{align}
&d_r=d_r^{(0)},
\nonumber\\
&d_R=d_R^{(0)},
\nonumber\\
&d_C=d_C^{(0)}-4\Lambda Cr(1-r),
\nonumber\\
&D_r=D_r^{(0)}+4\Lambda r^2(1-r)^2,
\nonumber\\
&D_R=D_R^{(0)}+4\Lambda C^2,
\nonumber\\
&D_C=D_C^{(0)} +4\Lambda C^2(1-2r)^2,
\nonumber\\
&D_{r,R}=D_{r,R}^{(0)}+4\Lambda Cr(1-r),
\nonumber\\
&D_{r,C}=D_{r,C}^{(0)}+4\Lambda Cr(1-r)(1-2r),
\nonumber\\
&D_{R,C}=D_{R,C}^{(0)} +4\Lambda C^2(1-2r).
\label{eq:alleqsonemonitored}
\end{align}
Here, the quantities with superscript $(0)$ refer to the expressions without monitoring,
given in Eq.~\eqref{eq:rrcddfree}.

From this, we find that subject to the same constraints as given in Eq.~\eqref{eq:prRc},
the joint distribution only explicitly depends on $r$,
\begin{equation}
P(r,R,C)=c_\Lambda\frac{1}{(1+8\Lambda r(1-r))^{3}}.
\label{eq:prrcmonitored}
\end{equation}
This allows us to obtain closed analytical expressions for the marginal distributions of these quantities, as well as the individual quantities $r_n$, and the concurrence $\mathcal{C}$. In particular, for the coefficient $r$, this gives the marginal distribution
\begin{align}
P(r)&=c_\Lambda\frac{r(1-r)}{(1+8\Lambda r(1-r) )^{3}}.
\label{eq:proneoftwo}
\end{align}
For the other quantities, highly nontrivial distributions arise as they depend on $r$ via the constraints, which is reflected by unwieldy formulas that we present in Appendix \ref{app:longexpressions}.

In Fig.~\ref{fig:oneoftwo}, these analytical results are compared to numerical results in a discretised stochastic time evolution with $\varepsilon=0.1$. In analogy to Fig.~\ref{fig:oneoftwofree}, the data is obtained from single quantum trajectories, but now with each unitary time step followed by a variable strength measurement as described in Sec.~
\ref{sec:twoqubitmonitoringprotocol}. We fix the microscopic measurement strength to $\lambda=0.1$ and $\lambda=\sqrt{0.05}\approx0.224$, corresponding to effective measurement strengths  $\Lambda=1$ and  $\Lambda=5$. As for the case without monitoring, we find good agreement, including for the probability coefficients $r$ and $R$ of the monitored and unmonitored qubits, which now follow different statistics.

We see that as the measurement strength increases, $P(r)$ again develops a bimodal shape peaked at $r=0$ and $r=1$, approaching the case of a hard projective measurement. In parallel, the distribution  $P(R)$ flattens out, slowly approaching the constant form of a single isolated qubit, while $P(C)$ become increasingly more confined to the regions of vanishing classical correlations ($C=0$),  and $P(\mathcal{C}^2)$ replicates this  as a trend towards vanishing quantum   correlations $\mathcal{C}^2=0$).

\subsection{Monitoring one of $q$ qubits}
\label{sec:oneofq}
\begin{figure}
  \centering
  \includegraphics[width=\columnwidth]{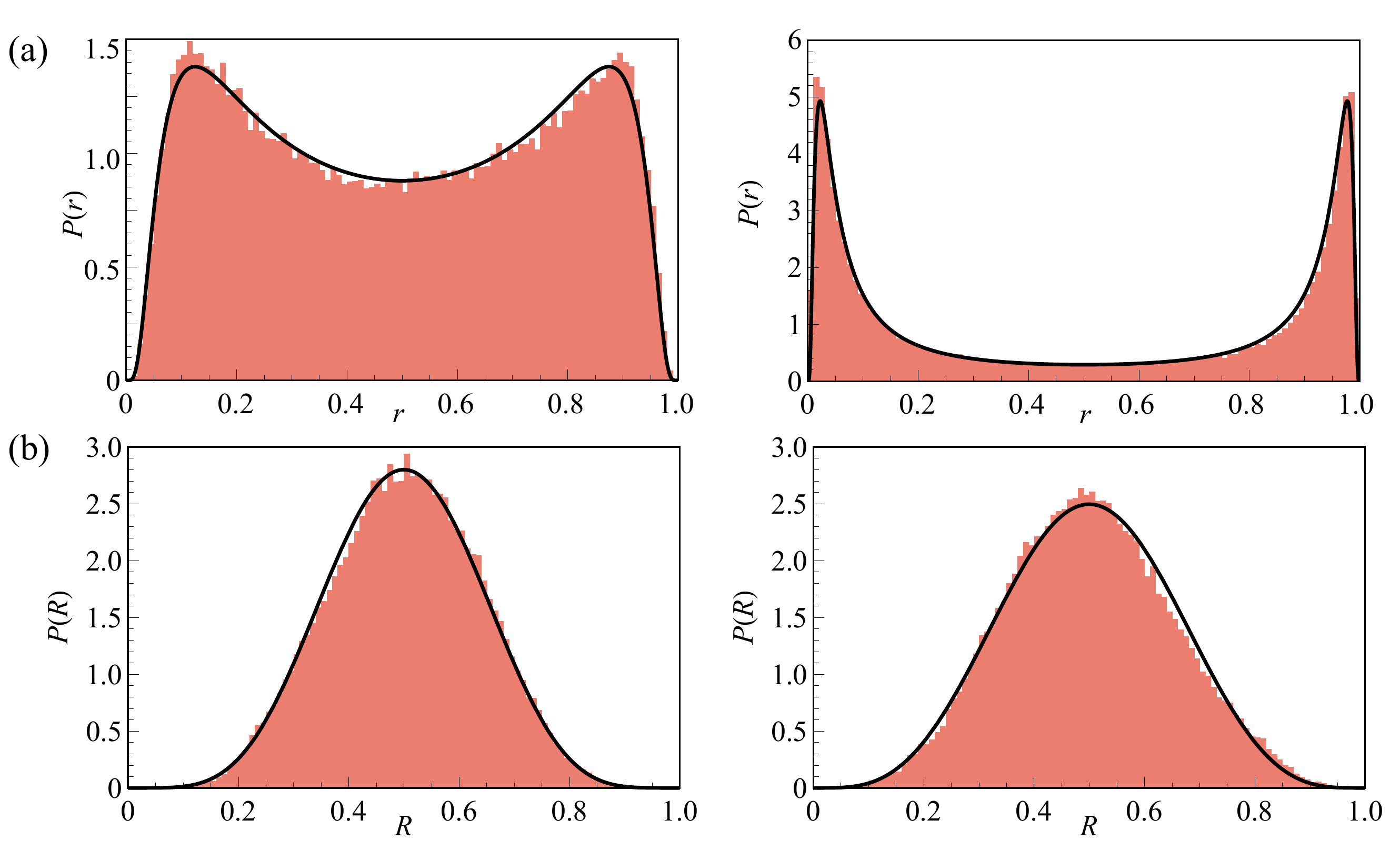}
  \caption{Monitoring one of four qubits. The panels show the probability densities of the probability coefficients $r$ of a designated monitored qubit and $R$ of one of the unmonitored qubits, for effective monitoring strength $\Lambda=1$ (left column) and $\Lambda=5$ (right column). The numerical data is obtained via the same process as underlying Fig.~\ref{fig:oneoftwo}, the analytical result for $P(r)$ is given by Eq.~\eqref{eq:proneofm},
while the result for $P(R)$ follows by combining this with the geometric constraints in the form of
Eq.~\eqref{eq:Reffectiveonefm}.}
\label{fig:oneoffour}
\end{figure}

For a larger collection of $q$ qubits, of which the first is monitored,
we obtain the drift, diffusion, and cross-correlation coefficients
\begin{align}
d_{r_n}&=\frac{1}{N}\left(\sum_{m\neq n}r_m-(N-1)r_n\right),
\\
D_{r_n}&=\frac{2}{N}r_n \sum_{m\neq n}r_m+4\Lambda r_n^2(r+x_n-1)^2,
\\
D_{r_n,r_m}&=-\frac{2}{N}r_nr_m+4\Lambda r_n r_m(r+x_n-1)(r+x_m-1).
\end{align}
Here $x_n=0$ for all indices $n\leq N/2$ in which the monitored qubit is in state $|0\rangle$ and $x_n=1$ when it is in state $|1\rangle$, while $r=\sum_{n| x_n=0}r_n$ is the probability coefficient of the monitored qubit, as already introduced in Eq.~\ref{eq:rshortcuts}.
These relations imply that the joint distribution of the probability coefficients is a function only of   $r$,
\begin{equation}
P(\{r_n\})=\frac{c_\Lambda}{(1+2N\Lambda r(1-r))^{N/2-1}}\delta(\sum_n r_n-1)
\end{equation}
with suitable normalization constant $c_\Lambda$,
capturing the constraints now via an over-parametrization in a delta function.
Indeed the equations for this coefficient continue to close, giving
\begin{align}
&d_r=\frac{1}{2}-r,\\
&D_r= \frac{2}{N}r(1-r)+4\Lambda r^2(1-r^2).
\end{align}
From this, we obtain the marginal distribution
\begin{equation}
P(r)=c_\Lambda\frac{[(1-r)r]^{(N/2)-1}}
{(1+2N\Lambda (1-r) r)^{(N/2)+1}}.
\label{eq:proneofm}
\end{equation}
This once more agrees very well with numerical simulations of the dynamics, as shown for a system of size $q=4$ in Fig. \ref{fig:oneoffour}.
Furthermore, we then obtain the statistics of other properties of the states by combining this with the geometric constraint, allowing us to express their probability distributions in terms of integrals.
In Fig. \ref{fig:oneoffour}, we illustrate this for the probability distribution of the summed probability coefficient $R$
of one of the non-monitored quantum bits. For this, the analytical prediction arises by writing
\begin{equation}
R_m=r\tilde{r}_0+(1-r)\tilde{r}_1,
\label{eq:Reffectiveonefm}
\end{equation}
where $r$ is distributed according to Eq. \eqref{eq:proneofm},
and $\tilde{r}_0$, $\tilde{r}_1$ independently follow the same distribution but with $\Lambda\to 0$, $N\to N/2$.
This again corresponds to a picture where the monitoring conditions the statistics of $r$, which then transfers to other statistical properties via geometric constraints.

\begin{figure}
  \centering
  \includegraphics[width=\columnwidth]{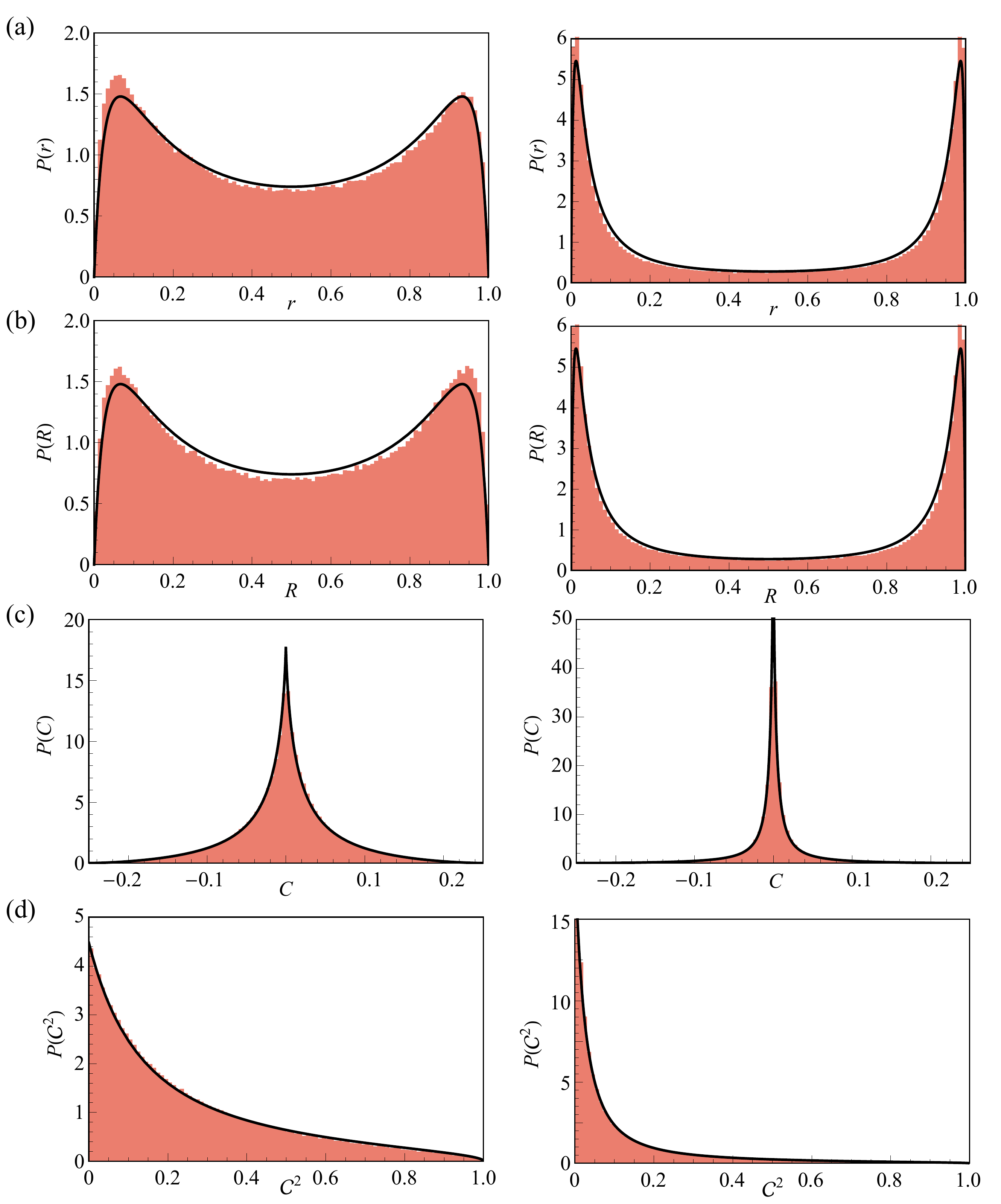}
  \caption{The numerical data shows the stationary quantum state statistics in a system of two qubits, where both are monitored with equal strength $\Lambda=\Lambda'=1$ (left column) or  $\Lambda=\Lambda'=5$ (right column).
In this case, the curves are not exact analytical predictions, but capture the phenomenological observation that these results bear similarities  to distributions in the system where just one of the two qubits is monitored, with suitable identifications of distributions and effective monitoring strengths as described in the text.}\label{fig:twooftwo}
\end{figure}

\section{Monitoring several qubits}
\label{sec:multiple}
Following the principles and procedure outlined in the previous section, we can also formulate the noisy dynamics of
systems in which multiple quantum bits are monitored. For the example of monitoring two quantum bits with effective strengths $\Lambda$ and $\Lambda'$, this amends the drift, diffusion, and cross-correlation coefficients
in Eq.~\eqref{eq:alleqsonemonitored}, to which we now refer with a superscript $(1)$,
to read
\begin{align}
&d_r=d_r^{(1)},
\nonumber\\
&d_R=d_R^{(1)},
\nonumber\\
&d_C=d_C^{(1)}-4\Lambda' CR(1-R),
\nonumber\\
&D_r=D_r^{(1)}+4\Lambda' C^2,
\nonumber\\
&D_R=D_R^{(1)}+4\Lambda' R^2(1-R)^2,
\nonumber\\
&D_C=D_C^{(1)}+4\Lambda' C^2(1-2R)^2,
\nonumber\\
&D_{r,R}=D_{r,R}^{(1)}+4\Lambda' CR(1-R),
\nonumber\\
&D_{r,C}=D_{r,C}^{(1)}+4\Lambda' C^2(1-2R),
\nonumber\\
&D_{R,C}=D_{R,R}^{(1)}+4\Lambda' CR(1-R)(1-2R) .
\label{eq:alleqstwomonitored}
\end{align}
While we then can formulate stationarity conditions that determine the joint probability distribution of the quantities at large times, these are not easily solved. Therefore, we here illustrate the resulting statistics based on numerical results.

In Fig.~\ref{fig:twooftwo}, we show the marginal distributions of $r$, $R$, $C$ and $\mathcal{C}^2$ for various combinations of $\Lambda=\Lambda'=1$ and
$\Lambda=\Lambda'=5$.
We then make the phenomenological observation that the distributions are similar to the distributions in a system with just a single monitored qubit, Sec.~\ref{sec:mononeoftwo}, with the following identifications.
The probability densities $P(r)=P(R)$ are close to Eq.~\eqref{eq:proneoftwo},  at $\Lambda_{\mathrm{eff}}=\Lambda=\Lambda'$. On the other hand, the probability densities for $C$ and $\mathcal{C}^2$ are close to Eqs.~\eqref{eq:longc} and \eqref{eq:longconc}, but with
$\Lambda_{\mathrm{eff}}=\Lambda+\Lambda'$.

\section{Application to a monitored kicked top}
\label{sec:kickedtop}
To round of this study, we consider monitoring in a deterministic system in which the noisy dynamics reflect the quantum signatures of chaos. For this, we choose the kicked top \cite{haake1987classical}, the dynamics of  angular momentum $(J_x,J_y,J_z)$ obtained from a Floquet operator
\begin{align}
F&=F_xF_y,\\
\nonumber
F_x&=\exp\left(-i\frac{k}{2j+1}J_x^2-i\beta_x J_x\right)
\nonumber\\
F_y&=\exp\left(-i\beta_y J_y\right)
\end{align}
that combines rotations by angles $\beta_x$ and $\beta_y$ with a torsion of strength $k$.
This system is a paradigm of complex quantum dynamics, inherited from a classical limit that turns chaotic at large enough torsion strength. This classical limit is attained when the angular quantum number $j$ becomes large.

We set
$\beta_x=0.8$, $\beta_y=2$, $k=8$, $j=15/2$ ($N=2j+1=16$), and slice each of the two factors $F_x$ and $F_y$ in the time evolution up into $n_T$ steps, between which we insert variable-strength measurement operations detecting whether the state is in the upper or lower hemisphere in the $J_z$ eigenbasis. This monitoring has just two possible outcomes with equal degeneracy, hence, with our choice of $j$, is equivalent to monitoring a designated qubit in a four-qubit system.
We then compare the statistics of a single quantum trajectory over $>10^5$ sliced time steps with the analytical results for this scenario in the stochastic model, where we set the effective monitoring strength to the natural value $\Lambda=\lambda^2n_T$ and keep $\lambda=0.1$ fixed.

\begin{figure}
  \centering
  \includegraphics[width=\columnwidth]{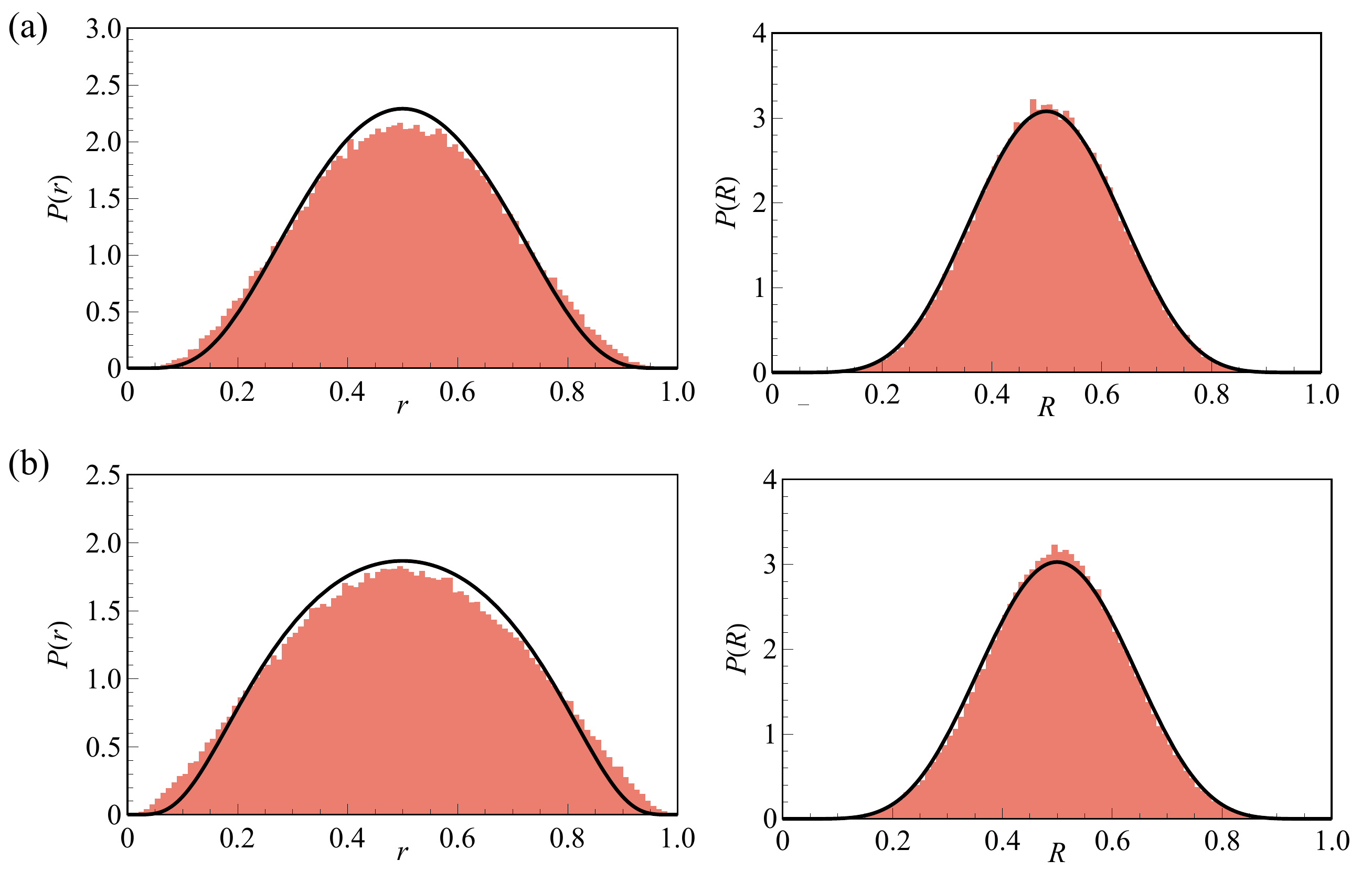}
  \caption{Comparison of numerical data for a monitored kicked top in the regime of weak monitoring, with
$\lambda=0.1$ and (a) $n_T=10$ or (b) $n_T=20$ (for further details see text),  to the analytical results in the equivalent setting of a noisy quantum circuit with (a) $\Lambda=0.1$ or (b) $\Lambda=0.2$.}\label{fig:ktop}
\end{figure}

We then identify two regimes.
As shown in Fig. \ref{fig:ktop}, for $n_T=10$ and $n_T=20$, where the corresponding $\Lambda=0.1,0.2$  is small, the results from the monitored kicked top are in acceptable agreement with the stochastic model, as captured in the analytical results of Sec.~\ref{sec:oneofq}. This is the regime where the alternating factors $F_x$ and $F_y$ mimic noise on time scales shorter than the monitoring time scale, which washes out any potential differences in the dynamical time scales of the two  factors.

\begin{figure}
  \centering
  \includegraphics[width=\columnwidth]{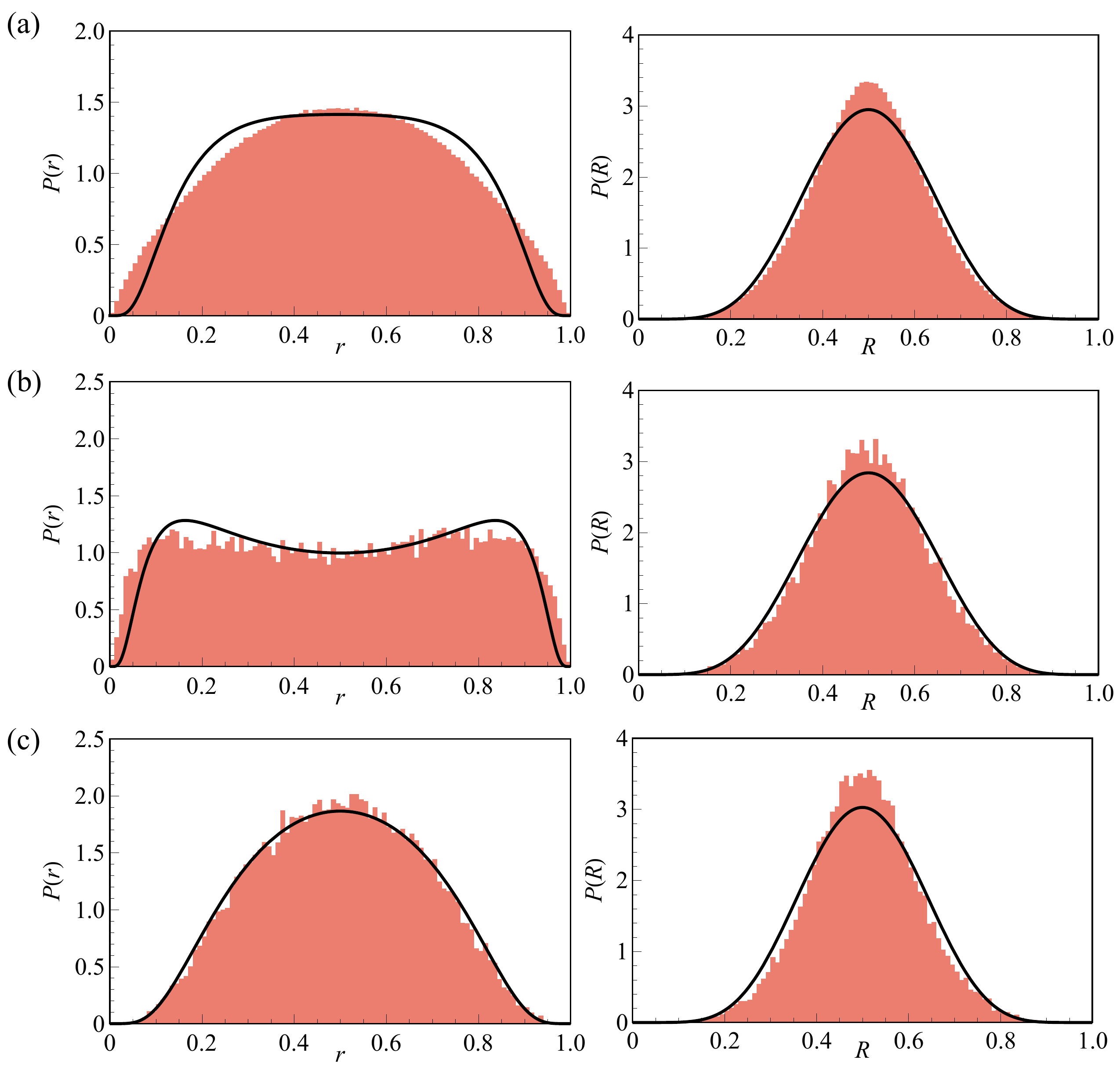}
  \caption{Comparison of numerical data for a monitored kicked top at stronger monitoring, with
$\lambda=0.1$ and $n_T=40$, recorded (a) at all time steps, (b) at the end of the rotation, and (c) at the end of the torsion,  to the analytical results in the equivalent setting of a noisy quantum circuit with (a) $\Lambda=0.4$, (b) $\Lambda=0.8$, and (c) $\Lambda=0.2$.}\label{fig:ktop2}
\end{figure}

However, as shown in Fig. \ref{fig:ktop2}, for the larger value $\Lambda=0.4$ the results from the monitored kicked top deviate from the stochastic model. Indeed, in this regime we obtain distinct statistics at the end of the rotation $F_y$ and at the end of the torsion $F_x$. Each of these statistics resembles, at least qualitatively, the data in the stochastic model at suitable effective strength $\Lambda_{\rm rot}\approx 0.2=2 \Lambda$ and $\Lambda_{\rm tor}\approx 0.8=\Lambda/2$, which can be interpreted as reflecting different levels of dynamical noise in $F_x$ and $F_y$ that are  resolved by the monitoring.
This demonstrates that the stochastic description developed in this work can be usefully applied to settings well outside the domain it has been initially defined, and then gives illuminating insights into the monitored dynamics of generic complex quantum systems.

\section{Conclusions}
\label{sec:conclusions}
In summary, in this work I formulated stochastic processes representing the noisy dynamics of monitored multi-qubit systems, and used these to describe the quasistationary statistics of the quantum state attained at long times. For the case of monitoring a single designated qubit in the system, this yields exact analytical descriptions in terms of the probability coefficients of the state. This leads to an understanding of the role of classical and quantum correlations in mediating the effects of monitoring. In a statistical interpretation, the monitoring conditions the state of the monitored qubit, which then translates into nontrivial statistics of the other qubits via geometric constraints. On the dynamical level, this indirect effect of the monitoring is mediated by correlations, such as the parameter embodied $C$ that I introduced to parameterise two-qubit states. These indirect effects also occur in more complicated setting where multiple qubits are monitored at different strengths, which we illustrated by numerical results complementing our analytical considerations.
Furthermore, the results obtained here also serve as a benchmark for monitoring in deterministic systems displaying the quantum signatures of chaos, as I illustrated for the quantum kicked top.

A specific open random-matrix problem arising from this work is to solve the case of multiple monitored qubits analytically. The framework described here allows to formulate the corresponding Fokker-Planck equations, but in these, the drift and diffusion coefficients then do no longer follow a hierarchical pattern, as is apparent when comparing the minimal case of two qubits, Eqs. \eqref{eq:alleqsonemonitored} and \eqref{eq:alleqstwomonitored}. A solution of the general case may still be possible by a suitable extension of the parametrisation $\eqref{eq:rRc0}$, which in itself appears to be a worthwhile task for its utility to extract statistical correlations and interpret the results physically and mathematical in terms of geometric constraints. For instance, it may well be possible to then formulate such constraints via a maximal-entropy principle, and identify the common statistical ground for the mathematically intriguing probability distributions that appear in this work.

Furthermore, a distinguishing feature of the noisy dynamics that we considered here is that these act globally on the complete system, mirroring the dynamics in a fully ergodic many-body quantum system. Indeed, we formulated the dynamics akin to Dyson's Brownian-motion process, which efficiently captures the statistics of such systems in a random-\emph{matrix} description, even though here this took the form of random \emph{states}. On the other hand, the measurement protocol employed in the monitoring extracts local information from the system. This work can therefore be extended in various directions. Amongst these, a particularly interesting case is that of quantum circuits built out of two-qubit gates that act locally. As mentioned in the introduction, these display a transition in the entanglement characteristics as the monitoring strength is increased, which was first realized in a stroboscopic setting, but also applies to continuously monitored noisy dynamics. Another variation is to consider the statistics for post selected dynamics, hence, the conditional probabilities obtained for a given, fixed, sequence of measurement outcomes. Conversely, the dynamics can be averaged incoherently over such outcomes, which captures decoherence in an open-system setting.
Furthermore, we can extend the considerations to physical components encompassing more than two levels, and again include deterministic systems in which the noisy dynamics reflect the quantum signatures of chaos, as we already did for a single such component in Sec.~\eqref{sec:kickedtop}. In these settings, one can
adopt the generalised monitoring protocol in Appendix \ref{app:b}, which can also be extended to describe monitoring  of different, possibly nonlocal, observables.
This raises the hope that the methods and findings presented in this paper inform the study of such systems, too.

\begin{acknowledgments}
This paper is dedicated to my dearly loved and respected Doktorvater, Fritz Haake.
The measurement process, random-matrix theory, and complex quantum systems displaying the signatures of underlying classical chaos are central themes in Fritzens Lebenswerk, which continues to inspire.
I also thank Tara Kalsi and Alessandro Romito for fruitful discussions that helped to shape the precise incarnation of these themes in this work.
\end{acknowledgments}

All numerical data in this work was directly obtained, processed, and plotted  in Mathematica, making use of its implementation of Gaussian unitary ensembles and the Wigner D function. This data is therefore completely represented by the figures.

\appendix

\section{General version of the measurement protocol}
\label{app:b}
For a general system of Hilbert-space dimension $N$,
we formulate the variable-strength measurement protocol in terms of a set of measurement parameters $\{\lambda_n\}_{n=1}^N$, each characterizing the strength by which the different basis states are probed.
This is achieved by controlled entanglement with an auxiliary system that has the same dimensionality, and then is measured.

\subsection{Entanglement step}
Initially, the joint system is in a separable state,
\begin{equation}
|\varphi\rangle=|\psi\rangle\otimes|0\rangle,
\end{equation}
where $|0\rangle$ is a suitable state of the auxiliary system.
The  entanglement step takes this into the state
\begin{equation}
|\varphi\rangle=\sum_{n=1}^N\psi_n|n\rangle\otimes|\lambda_n\rangle,
\end{equation}
where we set
\begin{equation}
|\lambda_n\rangle=\sum_{m=1}^N\sqrt{\frac{1-\lambda_n}{N}+\lambda_n \delta_{nm}}|m\rangle
\end{equation}
(note the dual role played by the index $n$ in this definition).
This systematically enhances one of the amplitudes of the auxiliary system in correspondence with a partner basis state of the system.

\subsection{Measurement step}
Measuring the auxiliary system then results in outcomes $M=1,2,\ldots,N$, which collapse the system state onto
\begin{align}
&|\psi_M\rangle=P_M^{-1/2}|\Psi_M\rangle,
\\
&|\Psi_M\rangle=\sum_{n=1}^N\sqrt{\frac{1-\lambda_n}{N}+\lambda_n \delta_{nM}}\psi_n|n\rangle,
\\
&P_M=\langle \Psi_M| \Psi_M\rangle=\lambda_M|\psi_M|^2+\sum_n\frac{1-\lambda_n}{N}|\psi_n|^2
\nonumber
\\
&\quad=\frac{1}{N}+\lambda_M|\psi_M|^2-
\sum_n\frac{\lambda_n}{N}|\psi_n|^2
,
\end{align}
where $P_M$ is the probability of  the measurement outcomes.
This sends the coefficients $r_{\tau_-,n}$ to
\begin{align}
r_{\tau_+,n}&=
r_{\tau_-,n} \frac{1-\lambda_n+N\lambda_n \delta_{nM}}{1+N\lambda_M r_{\tau_-,M}-
\sum_m \lambda_m r_{\tau_-,m}}
\\
&=
r_{\tau_-,n}+r_{\tau_-,n}
\frac{\sum_m \lambda_m (1-\delta_{mM})(r_{\tau_-,m}-\delta_{nm})}{1-
\sum_m \lambda_m (1-\delta_{mM})r_{\tau_-,m}}
\end{align}

Averaged over the measurement outcomes, we always have $\overline{r_{\tau_+,n}}=r_{\tau_-,n}$.

\subsection{Weak-measurement limit}

If the measurement strengths $\{\lambda_n\}$ are all small, we obtain a weak measurement, in which
the change of the
coefficients can be approximated as
\begin{align}
&dr_{\tau,n}\equiv r_{\tau_+,n}-r_{\tau_-,n}
\\
&=r_n \sum_m \lambda_m (1-N\delta_{mM})(r_m-\delta_{nm})
\nonumber\\
&\left(1+ \sum_m \lambda_m (1-N\delta_{mM})r_m\right)
+O(\lambda^3)
\end{align}

Averaged over the measurement outcomes, we then have
\begin{align}
&\overline{dr_{\tau,n}}=0,
\\
&\overline{dr_{\tau,n}dr_{\tau,m}}=
r_nr_m\Big[
N \sum _{l=1}^N (r_l- \delta _{nl}) (r_l- \delta _{ml}) \lambda_l^2
\nonumber
\\
&
-\sum _{l=1}^N (r_l- \delta _{nl})\lambda_l\sum _{l=1}^N (r_l- \delta _{ml})\lambda_l
  \Big].
\end{align}
where the latter expression also applies for $n=m$.

\section{Direct derivation of the joint distribution of $r$, $R$ and $C$}
\label{app:rRc}
Here we give a direct derivation of the uniform joint distribution $P(r,R,C)$, Eq.~\eqref{eq:prRc}, of a completely random two-qubit state parameterised according to Eq.~\eqref{eq:rRc}.
We  denote $(|\alpha|^2,|\beta|^2,|\gamma|^2,|\delta|^2)=(r_1,r_2,r_3,r_4)$ in accordance to Eq.~\eqref{eq:rn},
but momentarily lift the normalization of the state. This gives rise to the amended parametrization
\begin{align}
&r_1=\mathcal{N}[C+rR], \quad r_2=\mathcal{N}[r(1-R)-C],
\\ & r_3=\mathcal{N}[R(1-r)-C],\quad r_4=\mathcal{N}[(1-r)(1-R)+C],
\label{eq:rRcapp0}
\end{align}
corresponding to setting
\begin{align}
&r=\frac{r_1+r_2}{(r_1+r_2+r_3+r_4)},\\
&R=\frac{r_1+r_3}{(r_1+r_2+r_3+r_4)},\\
&C=\frac{r_1r_4-r_2r_3}{(r_1+r_2+r_3+r_4)^2},\\
&\mathcal{N}=r_1+r_2+r_3+r_4,
\end{align}
including the additional normalization parameter $\mathcal{N}$.

The Jacobian of the transformation is
\begin{equation}
J=\left|\det \frac{\partial(r_1,r_2,r_3,r_4)}{\partial(r,R,C,\mathcal{N})}\right|=\mathcal{N}^3.
\end{equation}

We then adopt any suitable distribution of the complex amplitudes $\psi_n$, such as the Gaussian
distribution
\begin{equation}
P(\{\psi_n\})=\exp(-\sum_{n=1}^4|\psi_n|^2)/\pi^4,
\end{equation}
that correctly reproduces the isotropic distribution of a normalized state when constrained to  $\mathcal{N}=1$.
Integrating out the complex phases, the distribution indeed only depends on the normalization parameter, such as here
\begin{equation}
P(\{r_n\})=\exp(-\sum_{n=1}^4r_n).
\end{equation}
Transformed to the new parameters, we then find \begin{equation}
P(r,R,C,\mathcal{N})=6\,\Theta(r,R,C)\times \frac{\mathcal{N}^3}{6}\exp(-\mathcal{N}),
\end{equation}
where
\begin{align}
\Theta(r,R,C) &=\Theta[C+rR] \times \Theta[-C +
   r (1 - R)]
\nonumber
\\
&{}\times \Theta[-C + (1 - r)R] \times\Theta[C + (1 - r) (1 -R)]
\nonumber
\\
&{}\times \Theta[r]\times \Theta[1-r]\times \Theta[R]\times \Theta[1-R]
\end{align}
embodies the constraints on the parameters in terms of the unit step function $\Theta[\cdot]$.
Importantly, these constraints are also independent of $\mathcal{N}$.
Therefore, for any value of the normalization parameter, the distribution $P(r,R,C)=6\Theta(r,R,C)$ is uniform, including for the normalized case  $\mathcal{N}=1$.

\begin{widetext}

\section{Explicit form of the analytical marginal distributions for monitoring one of two qubits}
\label{app:longexpressions}

Here, we collect the explicit analytical expressions of  marginal distributions  for  monitoring one of two qubits, discussed in subsection \ref{sec:mononeoftwo}. This is based on the joint distribution
\eqref{eq:prrcmonitored} for $r$, $R$, and $C$, subject to constraints in \eqref{eq:prRc}, as well as reinterpretations of these constraints in equivalent geometric terms.

First, we integrate out any two of these quantities subject to the stated constraints.
This entails the marginal distributions of the remaining quantity,
\begin{align}
P(r)&=c_\Lambda\frac{r(1-r)}{(1+8\Lambda r(1-r) )^{3}},
\\
P(R)&=
c_\Lambda\left(
\frac{8  R(1-R) }{1+8 \Lambda   R(1-R)}
+\frac{6}{\sqrt{2\Lambda}(2 \Lambda +1)^{3/2}} \Big[\asinh\left(\sqrt{2\Lambda }\right)
+(2 R-1) \atanh\left(\sqrt{\frac{2\Lambda }{2 \Lambda +1}} (1-2 R)\right)\Big]
\right)
\\
P(C)&=
c_\Lambda
\frac{4 \sqrt{1-4 \left| C\right| } \left(64 \Lambda ^2 \left| C\right| +\Lambda  (56 \left| C\right| +4)+5\right)}{(2 \Lambda +1) (8 \Lambda  \left| C\right| +1)}
\nonumber \\ &
+c_\Lambda\frac{2 \sqrt{2} \left(8 \Lambda  \left(32 \Lambda ^2+40 \Lambda +15\right) \left| C\right| +3\right) \atanh\left(\sqrt{\frac{\Lambda  (2-8 \left| C\right| )}{2 \Lambda +1}}\right)}{\sqrt{\Lambda } (2 \Lambda +1)^{3/2}}
-c_\Lambda(2 \Lambda +1)128\left| C\right| \atanh\left(\sqrt{1-4 \left| C\right| }\right).
\label{eq:longc}
\end{align}
In all these expressions, $c_\Lambda$
is a suitable normalization  constant.
Compared to the joint distribution \eqref{eq:prrcmonitored}, the marginally distributions are distinctively more nontrivial, and this is enforced by  the constraints in Eq.~\eqref{eq:prRc}.

This is useful as these constraints can be recovered in an equivalent picture, from which we can infer further statistics of the state.  In this picture,  the conditional states $|\tilde \psi_{0}\rangle$ and $|\tilde \psi_{1}\rangle$ of the second qubit, where the monitored qubit is in state $|0\rangle$ or $|1\rangle$, are uncorrelated, while their lengths are again conditioned by the probability coefficient $r$. Therefore,
\begin{align}
&\langle\tilde \psi_{0}|\tilde \psi_{0}\rangle =r,
&
\langle\tilde \psi_{1}|\tilde \psi_{1}\rangle =1-r,
&
|\langle\tilde \psi_{0}|\tilde \psi_{1}\rangle|^2 =r(1-r)x,
\end{align}
where $x$ is then uniformly distributed in $[0,1]$, independently of $x$.
In terms of this data, the squared concurrence takes the form
\begin{equation}
\mathcal{C}^2=4r(1-r)(1-x).
\end{equation}
From this, we can derive the corresponding probability distribution,
\begin{align}
P(\mathcal{C}^2)&=
\frac{c_\Lambda'}{(2 \mathcal{C}^2 \Lambda +1)^2}
\left(
4 ((6 \mathcal{C}^2+4) \Lambda +5) \sqrt{(1-\mathcal{C}^2) \Lambda  (2 \Lambda +1)}
+6 \sqrt{2} (2 \mathcal{C}^2\Lambda +1)^2 \atanh\left(\sqrt{\frac{(1-\mathcal{C}^2) 2\Lambda }{2 \Lambda +1}}\right)
\right)
\label{eq:longconc}
\end{align}
with $c_\Lambda'=c_\Lambda/[128 \sqrt{\Lambda } (2 \Lambda +1)^{5/2}]$.

Analogously we can interpret each of the components of the subvectors as random, and hence write $r_1=x'r$ where $0\leq x'\leq 1$, again with a uniform distribution. This gives
\begin{align}
P(r_n)=&
\frac{c_\Lambda'}{(1-8 \Lambda  (r_n-1) r_n)^2}
\Bigg[
8 \sqrt{\Lambda  (2 \Lambda +1)} (r_n-1)
[
-5 + 8 r_n + (32 r_n^2 (1 -\Lambda) - 4 (6 r_n + 1)) (1 -
     r_n) \Lambda]
\nonumber\\&
+6 \sqrt{2} \asinh\left(\sqrt{2\Lambda }\right) (1-8 \Lambda  (r_n-1) r_n)^2
+6 \sqrt{2} (1-8 \Lambda  (r_n-1) r_n)^2 \atanh\left(\frac{1-2 r_n}{\sqrt{\frac{1}{2 \Lambda }+1}}\right)
\Bigg].
\end{align}

These nontrivial distributions again combine the monitoring-conditioned statistics of the parameter $r$ with suitable $r$-dependent constraints.

\end{widetext}


\begin{thebibliography}{34}%
\makeatletter
\providecommand \@ifxundefined [1]{%
 \@ifx{#1\undefined}
}%
\providecommand \@ifnum [1]{%
 \ifnum #1\expandafter \@firstoftwo
 \else \expandafter \@secondoftwo
 \fi
}%
\providecommand \@ifx [1]{%
 \ifx #1\expandafter \@firstoftwo
 \else \expandafter \@secondoftwo
 \fi
}%
\providecommand \natexlab [1]{#1}%
\providecommand \enquote  [1]{``#1''}%
\providecommand \bibnamefont  [1]{#1}%
\providecommand \bibfnamefont [1]{#1}%
\providecommand \citenamefont [1]{#1}%
\providecommand \href@noop [0]{\@secondoftwo}%
\providecommand \href [0]{\begingroup \@sanitize@url \@href}%
\providecommand \@href[1]{\@@startlink{#1}\@@href}%
\providecommand \@@href[1]{\endgroup#1\@@endlink}%
\providecommand \@sanitize@url [0]{\catcode `\\12\catcode `\$12\catcode
  `\&12\catcode `\#12\catcode `\^12\catcode `\_12\catcode `\%12\relax}%
\providecommand \@@startlink[1]{}%
\providecommand \@@endlink[0]{}%
\providecommand \url  [0]{\begingroup\@sanitize@url \@url }%
\providecommand \@url [1]{\endgroup\@href {#1}{\urlprefix }}%
\providecommand \urlprefix  [0]{URL }%
\providecommand \Eprint [0]{\href }%
\providecommand \doibase [0]{https://doi.org/}%
\providecommand \selectlanguage [0]{\@gobble}%
\providecommand \bibinfo  [0]{\@secondoftwo}%
\providecommand \bibfield  [0]{\@secondoftwo}%
\providecommand \translation [1]{[#1]}%
\providecommand \BibitemOpen [0]{}%
\providecommand \bibitemStop [0]{}%
\providecommand \bibitemNoStop [0]{.\EOS\space}%
\providecommand \EOS [0]{\spacefactor3000\relax}%
\providecommand \BibitemShut  [1]{\csname bibitem#1\endcsname}%
\let\auto@bib@innerbib\@empty
\bibitem [{\citenamefont {von
  Neumann}(1938)}]{vonNeumann1938MathematicalTheory}%
  \BibitemOpen
  \bibfield  {author} {\bibinfo {author} {\bibfnamefont {J.}~\bibnamefont {von
  Neumann}},\ }\href@noop {} {\emph {\bibinfo {title} {{Mathematical Foundation
  of Quantum Theory}}}}\ (\bibinfo  {publisher} {Princeton University Press},\
  \bibinfo {address} {Princeton, NJ},\ \bibinfo {year} {1938})\BibitemShut
  {NoStop}%
\bibitem [{\citenamefont {Wiseman}\ and\ \citenamefont
  {Milburn}(2009)}]{Wiseman2009QuantumControl}%
  \BibitemOpen
  \bibfield  {author} {\bibinfo {author} {\bibfnamefont {H.~M.}\ \bibnamefont
  {Wiseman}}\ and\ \bibinfo {author} {\bibfnamefont {G.~J.}\ \bibnamefont
  {Milburn}},\ }\href {https://doi.org/10.1017/CBO9780511813948} {\emph
  {\bibinfo {title} {Quantum Measurement and Control}}}\ (\bibinfo  {publisher}
  {Cambridge University Press},\ \bibinfo {address} {Cambridge},\ \bibinfo
  {year} {2009})\BibitemShut {NoStop}%
\bibitem [{\citenamefont {Jacobs}(2014)}]{jacobs_2014}%
  \BibitemOpen
  \bibfield  {author} {\bibinfo {author} {\bibfnamefont {K.}~\bibnamefont
  {Jacobs}},\ }\href {https://doi.org/10.1017/CBO9781139179027} {\emph
  {\bibinfo {title} {Quantum Measurement Theory and its Applications}}}\
  (\bibinfo  {publisher} {Cambridge University Press},\ \bibinfo {year}
  {2014})\BibitemShut {NoStop}%
\bibitem [{\citenamefont {Haake}(1973)}]{haake1973statistical}%
  \BibitemOpen
  \bibfield  {author} {\bibinfo {author} {\bibfnamefont {F.}~\bibnamefont
  {Haake}},\ }\bibfield  {title} {\bibinfo {title} {Statistical treatment of
  open systems by generalized master equations},\ }in\ \href@noop {} {\emph
  {\bibinfo {booktitle} {Springer tracts in modern physics}}}\ (\bibinfo
  {publisher} {Springer},\ \bibinfo {year} {1973})\ pp.\ \bibinfo {pages}
  {98--168}\BibitemShut {NoStop}%
\bibitem [{\citenamefont {Gnutzmann}\ \emph {et~al.}(2021)\citenamefont
  {Gnutzmann}, \citenamefont {Guhr}, \citenamefont {Schomerus},\ and\
  \citenamefont {{\.{Z}}yczkowski}}]{Gnutzmann_2021}%
  \BibitemOpen
  \bibfield  {author} {\bibinfo {author} {\bibfnamefont {S.}~\bibnamefont
  {Gnutzmann}}, \bibinfo {author} {\bibfnamefont {T.}~\bibnamefont {Guhr}},
  \bibinfo {author} {\bibfnamefont {H.}~\bibnamefont {Schomerus}},\ and\
  \bibinfo {author} {\bibfnamefont {K.}~\bibnamefont {{\.{Z}}yczkowski}},\
  }\bibfield  {title} {\bibinfo {title} {Special issue in honour of the life
  and work of {Fritz} {Haake}},\ }\href
  {https://doi.org/10.1088/1751-8121/abe167} {\bibfield  {journal} {\bibinfo
  {journal} {J. Phys. A}\ }\textbf {\bibinfo {volume} {54}},\ \bibinfo {pages}
  {130301} (\bibinfo {year} {2021})}\BibitemShut {NoStop}%
\bibitem [{\citenamefont {Carmichael}(2009)}]{carmichael2009open}%
  \BibitemOpen
  \bibfield  {author} {\bibinfo {author} {\bibfnamefont {H.}~\bibnamefont
  {Carmichael}},\ }\href {https://books.google.co.uk/books?id=uor\_CAAAQBAJ}
  {\emph {\bibinfo {title} {An Open Systems Approach to Quantum Optics:
  Lectures Presented at the Universit{\'e} Libre de Bruxelles, October 28 to
  November 4, 1991}}},\ Lecture Notes in Physics Monographs\ (\bibinfo
  {publisher} {Springer Berlin Heidelberg},\ \bibinfo {year}
  {2009})\BibitemShut {NoStop}%
\bibitem [{\citenamefont {Chan}\ \emph {et~al.}(2019)\citenamefont {Chan},
  \citenamefont {Nandkishore}, \citenamefont {Pretko},\ and\ \citenamefont
  {Smith}}]{Chan2019Unitary-projectiveDynamics}%
  \BibitemOpen
  \bibfield  {author} {\bibinfo {author} {\bibfnamefont {A.}~\bibnamefont
  {Chan}}, \bibinfo {author} {\bibfnamefont {R.~M.}\ \bibnamefont
  {Nandkishore}}, \bibinfo {author} {\bibfnamefont {M.}~\bibnamefont
  {Pretko}},\ and\ \bibinfo {author} {\bibfnamefont {G.}~\bibnamefont
  {Smith}},\ }\bibfield  {title} {\bibinfo {title} {{Unitary-projective
  entanglement dynamics}},\ }\href {https://doi.org/10.1103/PhysRevB.99.224307}
  {\bibfield  {journal} {\bibinfo  {journal} {Phys. Rev. B}\ }\textbf {\bibinfo
  {volume} {99}},\ \bibinfo {pages} {224307} (\bibinfo {year}
  {2019})}\BibitemShut {NoStop}%
\bibitem [{\citenamefont {Skinner}\ \emph {et~al.}(2019)\citenamefont
  {Skinner}, \citenamefont {Ruhman},\ and\ \citenamefont
  {Nahum}}]{Skinner2019Measurement-InducedEntanglement}%
  \BibitemOpen
  \bibfield  {author} {\bibinfo {author} {\bibfnamefont {B.}~\bibnamefont
  {Skinner}}, \bibinfo {author} {\bibfnamefont {J.}~\bibnamefont {Ruhman}},\
  and\ \bibinfo {author} {\bibfnamefont {A.}~\bibnamefont {Nahum}},\ }\bibfield
   {title} {\bibinfo {title} {Measurement-induced phase transitions in the
  dynamics of entanglement},\ }\href
  {https://doi.org/10.1103/PhysRevX.9.031009} {\bibfield  {journal} {\bibinfo
  {journal} {Phys. Rev. X}\ }\textbf {\bibinfo {volume} {9}},\ \bibinfo {pages}
  {031009} (\bibinfo {year} {2019})}\BibitemShut {NoStop}%
\bibitem [{\citenamefont {Li}\ \emph {et~al.}(2018)\citenamefont {Li},
  \citenamefont {Chen},\ and\ \citenamefont
  {Fisher}}]{Li2018QuantumTransition}%
  \BibitemOpen
  \bibfield  {author} {\bibinfo {author} {\bibfnamefont {Y.}~\bibnamefont
  {Li}}, \bibinfo {author} {\bibfnamefont {X.}~\bibnamefont {Chen}},\ and\
  \bibinfo {author} {\bibfnamefont {M.~P.~A.}\ \bibnamefont {Fisher}},\
  }\bibfield  {title} {\bibinfo {title} {{Quantum {Zeno} effect and the
  many-body entanglement transition}},\ }\href
  {https://doi.org/10.1103/PhysRevB.98.205136} {\bibfield  {journal} {\bibinfo
  {journal} {Phys. Rev. B}\ }\textbf {\bibinfo {volume} {98}},\ \bibinfo
  {pages} {205136} (\bibinfo {year} {2018})}\BibitemShut {NoStop}%
\bibitem [{\citenamefont {Li}\ \emph {et~al.}(2019)\citenamefont {Li},
  \citenamefont {Chen},\ and\ \citenamefont
  {Fisher}}]{Li2019Measurement-drivenCircuits}%
  \BibitemOpen
  \bibfield  {author} {\bibinfo {author} {\bibfnamefont {Y.}~\bibnamefont
  {Li}}, \bibinfo {author} {\bibfnamefont {X.}~\bibnamefont {Chen}},\ and\
  \bibinfo {author} {\bibfnamefont {M.~P.~A.}\ \bibnamefont {Fisher}},\
  }\bibfield  {title} {\bibinfo {title} {{Measurement-driven entanglement
  transition in hybrid quantum circuits}},\ }\href
  {https://doi.org/10.1103/PhysRevB.100.134306} {\bibfield  {journal} {\bibinfo
   {journal} {Phys. Rev. B}\ }\textbf {\bibinfo {volume} {100}},\ \bibinfo
  {pages} {134306} (\bibinfo {year} {2019})}\BibitemShut {NoStop}%
\bibitem [{\citenamefont {Li}\ and\ \citenamefont {Fisher}(2021)}]{Li2021}%
  \BibitemOpen
  \bibfield  {author} {\bibinfo {author} {\bibfnamefont {Y.}~\bibnamefont
  {Li}}\ and\ \bibinfo {author} {\bibfnamefont {M.~P.~A.}\ \bibnamefont
  {Fisher}},\ }\bibfield  {title} {\bibinfo {title} {Statistical mechanics of
  quantum error correcting codes},\ }\href
  {https://doi.org/10.1103/PhysRevB.103.104306} {\bibfield  {journal} {\bibinfo
   {journal} {Phys. Rev. B}\ }\textbf {\bibinfo {volume} {103}},\ \bibinfo
  {pages} {104306} (\bibinfo {year} {2021})}\BibitemShut {NoStop}%
\bibitem [{\citenamefont {Gullans}\ and\ \citenamefont
  {Huse}(2020{\natexlab{a}})}]{Gullans2019purification}%
  \BibitemOpen
  \bibfield  {author} {\bibinfo {author} {\bibfnamefont {M.~J.}\ \bibnamefont
  {Gullans}}\ and\ \bibinfo {author} {\bibfnamefont {D.~A.}\ \bibnamefont
  {Huse}},\ }\bibfield  {title} {\bibinfo {title} {Dynamical purification phase
  transition induced by quantum measurements},\ }\href
  {https://doi.org/10.1103/PhysRevX.10.041020} {\bibfield  {journal} {\bibinfo
  {journal} {Phys. Rev. X}\ }\textbf {\bibinfo {volume} {10}},\ \bibinfo
  {pages} {041020} (\bibinfo {year} {2020}{\natexlab{a}})}\BibitemShut
  {NoStop}%
\bibitem [{\citenamefont {Gullans}\ and\ \citenamefont
  {Huse}(2020{\natexlab{b}})}]{Gullans2020}%
  \BibitemOpen
  \bibfield  {author} {\bibinfo {author} {\bibfnamefont {M.~J.}\ \bibnamefont
  {Gullans}}\ and\ \bibinfo {author} {\bibfnamefont {D.~A.}\ \bibnamefont
  {Huse}},\ }\bibfield  {title} {\bibinfo {title} {Scalable probes of
  measurement-induced criticality},\ }\href
  {https://doi.org/10.1103/PhysRevLett.125.070606} {\bibfield  {journal}
  {\bibinfo  {journal} {Phys. Rev. Lett.}\ }\textbf {\bibinfo {volume} {125}},\
  \bibinfo {pages} {070606} (\bibinfo {year} {2020}{\natexlab{b}})}\BibitemShut
  {NoStop}%
\bibitem [{\citenamefont {Zabalo}\ \emph {et~al.}(2020)\citenamefont {Zabalo},
  \citenamefont {Gullans}, \citenamefont {Wilson}, \citenamefont
  {Gopalakrishnan}, \citenamefont {Huse},\ and\ \citenamefont
  {Pixley}}]{Zabalo2020CriticalCircuits}%
  \BibitemOpen
  \bibfield  {author} {\bibinfo {author} {\bibfnamefont {A.}~\bibnamefont
  {Zabalo}}, \bibinfo {author} {\bibfnamefont {M.~J.}\ \bibnamefont {Gullans}},
  \bibinfo {author} {\bibfnamefont {J.~H.}\ \bibnamefont {Wilson}}, \bibinfo
  {author} {\bibfnamefont {S.}~\bibnamefont {Gopalakrishnan}}, \bibinfo
  {author} {\bibfnamefont {D.~A.}\ \bibnamefont {Huse}},\ and\ \bibinfo
  {author} {\bibfnamefont {J.~H.}\ \bibnamefont {Pixley}},\ }\bibfield  {title}
  {\bibinfo {title} {{Critical properties of the measurement-induced transition
  in random quantum circuits}},\ }\href
  {https://doi.org/10.1103/PhysRevB.101.060301} {\bibfield  {journal} {\bibinfo
   {journal} {Phys. Rev. B}\ }\textbf {\bibinfo {volume} {101}},\ \bibinfo
  {pages} {060301} (\bibinfo {year} {2020})}\BibitemShut {NoStop}%
\bibitem [{\citenamefont {Lunt}\ \emph {et~al.}(2021)\citenamefont {Lunt},
  \citenamefont {Szyniszewski},\ and\ \citenamefont
  {Pal}}]{PhysRevB.104.155111}%
  \BibitemOpen
  \bibfield  {author} {\bibinfo {author} {\bibfnamefont {O.}~\bibnamefont
  {Lunt}}, \bibinfo {author} {\bibfnamefont {M.}~\bibnamefont {Szyniszewski}},\
  and\ \bibinfo {author} {\bibfnamefont {A.}~\bibnamefont {Pal}},\ }\bibfield
  {title} {\bibinfo {title} {Measurement-induced criticality and entanglement
  clusters: A study of one-dimensional and two-dimensional {Clifford}
  circuits},\ }\href {https://doi.org/10.1103/PhysRevB.104.155111} {\bibfield
  {journal} {\bibinfo  {journal} {Phys. Rev. B}\ }\textbf {\bibinfo {volume}
  {104}},\ \bibinfo {pages} {155111} (\bibinfo {year} {2021})}\BibitemShut
  {NoStop}%
\bibitem [{\citenamefont {Zabalo}\ \emph {et~al.}(2022)\citenamefont {Zabalo},
  \citenamefont {Gullans}, \citenamefont {Wilson}, \citenamefont {Vasseur},
  \citenamefont {Ludwig}, \citenamefont {Gopalakrishnan}, \citenamefont
  {Huse},\ and\ \citenamefont {Pixley}}]{Zabalo2022}%
  \BibitemOpen
  \bibfield  {author} {\bibinfo {author} {\bibfnamefont {A.}~\bibnamefont
  {Zabalo}}, \bibinfo {author} {\bibfnamefont {M.~J.}\ \bibnamefont {Gullans}},
  \bibinfo {author} {\bibfnamefont {J.~H.}\ \bibnamefont {Wilson}}, \bibinfo
  {author} {\bibfnamefont {R.}~\bibnamefont {Vasseur}}, \bibinfo {author}
  {\bibfnamefont {A.~W.~W.}\ \bibnamefont {Ludwig}}, \bibinfo {author}
  {\bibfnamefont {S.}~\bibnamefont {Gopalakrishnan}}, \bibinfo {author}
  {\bibfnamefont {D.~A.}\ \bibnamefont {Huse}},\ and\ \bibinfo {author}
  {\bibfnamefont {J.~H.}\ \bibnamefont {Pixley}},\ }\bibfield  {title}
  {\bibinfo {title} {Operator scaling dimensions and multifractality at
  measurement-induced transitions},\ }\href
  {https://doi.org/10.1103/PhysRevLett.128.050602} {\bibfield  {journal}
  {\bibinfo  {journal} {Phys. Rev. Lett.}\ }\textbf {\bibinfo {volume} {128}},\
  \bibinfo {pages} {050602} (\bibinfo {year} {2022})}\BibitemShut {NoStop}%
\bibitem [{\citenamefont {Li}\ \emph {et~al.}(2021)\citenamefont {Li},
  \citenamefont {Chen}, \citenamefont {Ludwig},\ and\ \citenamefont
  {Fisher}}]{Li2020Conformal}%
  \BibitemOpen
  \bibfield  {author} {\bibinfo {author} {\bibfnamefont {Y.}~\bibnamefont
  {Li}}, \bibinfo {author} {\bibfnamefont {X.}~\bibnamefont {Chen}}, \bibinfo
  {author} {\bibfnamefont {A.~W.~W.}\ \bibnamefont {Ludwig}},\ and\ \bibinfo
  {author} {\bibfnamefont {M.~P.~A.}\ \bibnamefont {Fisher}},\ }\bibfield
  {title} {\bibinfo {title} {Conformal invariance and quantum nonlocality in
  critical hybrid circuits},\ }\href
  {https://doi.org/10.1103/PhysRevB.104.104305} {\bibfield  {journal} {\bibinfo
   {journal} {Phys. Rev. B}\ }\textbf {\bibinfo {volume} {104}},\ \bibinfo
  {pages} {104305} (\bibinfo {year} {2021})}\BibitemShut {NoStop}%
\bibitem [{\citenamefont {Bao}\ \emph {et~al.}(2020)\citenamefont {Bao},
  \citenamefont {Choi},\ and\ \citenamefont {Altman}}]{Bao2020}%
  \BibitemOpen
  \bibfield  {author} {\bibinfo {author} {\bibfnamefont {Y.}~\bibnamefont
  {Bao}}, \bibinfo {author} {\bibfnamefont {S.}~\bibnamefont {Choi}},\ and\
  \bibinfo {author} {\bibfnamefont {E.}~\bibnamefont {Altman}},\ }\bibfield
  {title} {\bibinfo {title} {Theory of the phase transition in random unitary
  circuits with measurements},\ }\href
  {https://doi.org/10.1103/PhysRevB.101.104301} {\bibfield  {journal} {\bibinfo
   {journal} {Phys. Rev. B}\ }\textbf {\bibinfo {volume} {101}},\ \bibinfo
  {pages} {104301} (\bibinfo {year} {2020})}\BibitemShut {NoStop}%
\bibitem [{\citenamefont {Jian}\ \emph {et~al.}(2020)\citenamefont {Jian},
  \citenamefont {You}, \citenamefont {Vasseur},\ and\ \citenamefont
  {Ludwig}}]{Jian2020}%
  \BibitemOpen
  \bibfield  {author} {\bibinfo {author} {\bibfnamefont {C.-M.}\ \bibnamefont
  {Jian}}, \bibinfo {author} {\bibfnamefont {Y.-Z.}\ \bibnamefont {You}},
  \bibinfo {author} {\bibfnamefont {R.}~\bibnamefont {Vasseur}},\ and\ \bibinfo
  {author} {\bibfnamefont {A.~W.~W.}\ \bibnamefont {Ludwig}},\ }\bibfield
  {title} {\bibinfo {title} {Measurement-induced criticality in random quantum
  circuits},\ }\href {https://doi.org/10.1103/PhysRevB.101.104302} {\bibfield
  {journal} {\bibinfo  {journal} {Phys. Rev. B}\ }\textbf {\bibinfo {volume}
  {101}},\ \bibinfo {pages} {104302} (\bibinfo {year} {2020})}\BibitemShut
  {NoStop}%
\bibitem [{\citenamefont {Fan}\ \emph {et~al.}(2021)\citenamefont {Fan},
  \citenamefont {Vijay}, \citenamefont {Vishwanath},\ and\ \citenamefont
  {You}}]{Fan2021}%
  \BibitemOpen
  \bibfield  {author} {\bibinfo {author} {\bibfnamefont {R.}~\bibnamefont
  {Fan}}, \bibinfo {author} {\bibfnamefont {S.}~\bibnamefont {Vijay}}, \bibinfo
  {author} {\bibfnamefont {A.}~\bibnamefont {Vishwanath}},\ and\ \bibinfo
  {author} {\bibfnamefont {Y.-Z.}\ \bibnamefont {You}},\ }\bibfield  {title}
  {\bibinfo {title} {Self-organized error correction in random unitary circuits
  with measurement},\ }\href {https://doi.org/10.1103/PhysRevB.103.174309}
  {\bibfield  {journal} {\bibinfo  {journal} {Phys. Rev. B}\ }\textbf {\bibinfo
  {volume} {103}},\ \bibinfo {pages} {174309} (\bibinfo {year}
  {2021})}\BibitemShut {NoStop}%
\bibitem [{\citenamefont {Bao}\ \emph {et~al.}(2021)\citenamefont {Bao},
  \citenamefont {Choi},\ and\ \citenamefont {Altman}}]{Bao2021}%
  \BibitemOpen
  \bibfield  {author} {\bibinfo {author} {\bibfnamefont {Y.}~\bibnamefont
  {Bao}}, \bibinfo {author} {\bibfnamefont {S.}~\bibnamefont {Choi}},\ and\
  \bibinfo {author} {\bibfnamefont {E.}~\bibnamefont {Altman}},\ }\bibfield
  {title} {\bibinfo {title} {Symmetry enriched phases of quantum circuits},\
  }\href {https://doi.org/https://doi.org/10.1016/j.aop.2021.168618} {\bibfield
   {journal} {\bibinfo  {journal} {Ann. Phys.}\ }\textbf {\bibinfo {volume}
  {435}},\ \bibinfo {pages} {168618} (\bibinfo {year} {2021})},\ \bibinfo
  {note} {special issue on Philip W. Anderson}\BibitemShut {NoStop}%
\bibitem [{\citenamefont {Bera}\ and\ \citenamefont
  {Singha~Roy}(2020)}]{Bera2020}%
  \BibitemOpen
  \bibfield  {author} {\bibinfo {author} {\bibfnamefont {A.}~\bibnamefont
  {Bera}}\ and\ \bibinfo {author} {\bibfnamefont {S.}~\bibnamefont
  {Singha~Roy}},\ }\bibfield  {title} {\bibinfo {title} {Growth of genuine
  multipartite entanglement in random unitary circuits},\ }\href
  {https://doi.org/10.1103/PhysRevA.102.062431} {\bibfield  {journal} {\bibinfo
   {journal} {Phys. Rev. A}\ }\textbf {\bibinfo {volume} {102}},\ \bibinfo
  {pages} {062431} (\bibinfo {year} {2020})}\BibitemShut {NoStop}%
\bibitem [{\citenamefont {Sang}\ and\ \citenamefont
  {Hsieh}(2021)}]{Sang2021measurement}%
  \BibitemOpen
  \bibfield  {author} {\bibinfo {author} {\bibfnamefont {S.}~\bibnamefont
  {Sang}}\ and\ \bibinfo {author} {\bibfnamefont {T.~H.}\ \bibnamefont
  {Hsieh}},\ }\bibfield  {title} {\bibinfo {title} {Measurement-protected
  quantum phases},\ }\href {https://doi.org/10.1103/PhysRevResearch.3.023200}
  {\bibfield  {journal} {\bibinfo  {journal} {Phys. Rev. Research}\ }\textbf
  {\bibinfo {volume} {3}},\ \bibinfo {pages} {023200} (\bibinfo {year}
  {2021})}\BibitemShut {NoStop}%
\bibitem [{\citenamefont {Zhang}\ \emph {et~al.}(2020)\citenamefont {Zhang},
  \citenamefont {Reyes}, \citenamefont {Kourtis}, \citenamefont {Chamon},
  \citenamefont {Mucciolo},\ and\ \citenamefont {Ruckenstein}}]{Zhang2020}%
  \BibitemOpen
  \bibfield  {author} {\bibinfo {author} {\bibfnamefont {L.}~\bibnamefont
  {Zhang}}, \bibinfo {author} {\bibfnamefont {J.~A.}\ \bibnamefont {Reyes}},
  \bibinfo {author} {\bibfnamefont {S.}~\bibnamefont {Kourtis}}, \bibinfo
  {author} {\bibfnamefont {C.}~\bibnamefont {Chamon}}, \bibinfo {author}
  {\bibfnamefont {E.~R.}\ \bibnamefont {Mucciolo}},\ and\ \bibinfo {author}
  {\bibfnamefont {A.~E.}\ \bibnamefont {Ruckenstein}},\ }\bibfield  {title}
  {\bibinfo {title} {Nonuniversal entanglement level statistics in
  projection-driven quantum circuits},\ }\href
  {https://doi.org/10.1103/PhysRevB.101.235104} {\bibfield  {journal} {\bibinfo
   {journal} {Phys. Rev. B}\ }\textbf {\bibinfo {volume} {101}},\ \bibinfo
  {pages} {235104} (\bibinfo {year} {2020})}\BibitemShut {NoStop}%
\bibitem [{\citenamefont {Choi}\ \emph {et~al.}(2020)\citenamefont {Choi},
  \citenamefont {Bao}, \citenamefont {Qi},\ and\ \citenamefont
  {Altman}}]{Choi2020}%
  \BibitemOpen
  \bibfield  {author} {\bibinfo {author} {\bibfnamefont {S.}~\bibnamefont
  {Choi}}, \bibinfo {author} {\bibfnamefont {Y.}~\bibnamefont {Bao}}, \bibinfo
  {author} {\bibfnamefont {X.-L.}\ \bibnamefont {Qi}},\ and\ \bibinfo {author}
  {\bibfnamefont {E.}~\bibnamefont {Altman}},\ }\bibfield  {title} {\bibinfo
  {title} {Quantum error correction in scrambling dynamics and
  measurement-induced phase transition},\ }\href
  {https://doi.org/10.1103/PhysRevLett.125.030505} {\bibfield  {journal}
  {\bibinfo  {journal} {Phys. Rev. Lett.}\ }\textbf {\bibinfo {volume} {125}},\
  \bibinfo {pages} {030505} (\bibinfo {year} {2020})}\BibitemShut {NoStop}%
\bibitem [{\citenamefont {Nahum}\ \emph {et~al.}(2021)\citenamefont {Nahum},
  \citenamefont {Roy}, \citenamefont {Skinner},\ and\ \citenamefont
  {Ruhman}}]{Nahum2021}%
  \BibitemOpen
  \bibfield  {author} {\bibinfo {author} {\bibfnamefont {A.}~\bibnamefont
  {Nahum}}, \bibinfo {author} {\bibfnamefont {S.}~\bibnamefont {Roy}}, \bibinfo
  {author} {\bibfnamefont {B.}~\bibnamefont {Skinner}},\ and\ \bibinfo {author}
  {\bibfnamefont {J.}~\bibnamefont {Ruhman}},\ }\bibfield  {title} {\bibinfo
  {title} {Measurement and entanglement phase transitions in all-to-all quantum
  circuits, on quantum trees, and in {Landau-Ginsburg} theory},\ }\href
  {https://doi.org/10.1103/PRXQuantum.2.010352} {\bibfield  {journal} {\bibinfo
   {journal} {PRX Quantum}\ }\textbf {\bibinfo {volume} {2}},\ \bibinfo {pages}
  {010352} (\bibinfo {year} {2021})}\BibitemShut {NoStop}%
\bibitem [{\citenamefont {Rossini}\ and\ \citenamefont
  {Vicari}(2020)}]{Rossini2020}%
  \BibitemOpen
  \bibfield  {author} {\bibinfo {author} {\bibfnamefont {D.}~\bibnamefont
  {Rossini}}\ and\ \bibinfo {author} {\bibfnamefont {E.}~\bibnamefont
  {Vicari}},\ }\bibfield  {title} {\bibinfo {title} {Measurement-induced
  dynamics of many-body systems at quantum criticality},\ }\href
  {https://doi.org/10.1103/PhysRevB.102.035119} {\bibfield  {journal} {\bibinfo
   {journal} {Phys. Rev. B}\ }\textbf {\bibinfo {volume} {102}},\ \bibinfo
  {pages} {035119} (\bibinfo {year} {2020})}\BibitemShut {NoStop}%
\bibitem [{\citenamefont {Iaconis}\ \emph {et~al.}(2020)\citenamefont
  {Iaconis}, \citenamefont {Lucas},\ and\ \citenamefont {Chen}}]{Iaconis2020}%
  \BibitemOpen
  \bibfield  {author} {\bibinfo {author} {\bibfnamefont {J.}~\bibnamefont
  {Iaconis}}, \bibinfo {author} {\bibfnamefont {A.}~\bibnamefont {Lucas}},\
  and\ \bibinfo {author} {\bibfnamefont {X.}~\bibnamefont {Chen}},\ }\bibfield
  {title} {\bibinfo {title} {Measurement-induced phase transitions in quantum
  automaton circuits},\ }\href {https://doi.org/10.1103/PhysRevB.102.224311}
  {\bibfield  {journal} {\bibinfo  {journal} {Phys. Rev. B}\ }\textbf {\bibinfo
  {volume} {102}},\ \bibinfo {pages} {224311} (\bibinfo {year}
  {2020})}\BibitemShut {NoStop}%
\bibitem [{\citenamefont {Kalsi}\ \emph {et~al.}(2022)\citenamefont {Kalsi},
  \citenamefont {Romito},\ and\ \citenamefont
  {Schomerus}}]{Kalsi2022Threefold}%
  \BibitemOpen
  \bibfield  {author} {\bibinfo {author} {\bibfnamefont {T.}~\bibnamefont
  {Kalsi}}, \bibinfo {author} {\bibfnamefont {A.}~\bibnamefont {Romito}},\ and\
  \bibinfo {author} {\bibfnamefont {H.}~\bibnamefont {Schomerus}},\ }\href@noop
  {} {\bibinfo {title} {Three-fold way of entanglement dynamics in monitored
  quantum circuits}} (\bibinfo {year} {2022}),\ \Eprint
  {https://arxiv.org/abs/2201.12259} {arXiv:2201.12259} \BibitemShut {NoStop}%
\bibitem [{\citenamefont {Szyniszewski}\ \emph {et~al.}(2019)\citenamefont
  {Szyniszewski}, \citenamefont {Romito},\ and\ \citenamefont
  {Schomerus}}]{Szyniszewski2019EntanglementMeasurements}%
  \BibitemOpen
  \bibfield  {author} {\bibinfo {author} {\bibfnamefont {M.}~\bibnamefont
  {Szyniszewski}}, \bibinfo {author} {\bibfnamefont {A.}~\bibnamefont
  {Romito}},\ and\ \bibinfo {author} {\bibfnamefont {H.}~\bibnamefont
  {Schomerus}},\ }\bibfield  {title} {\bibinfo {title} {{Entanglement
  transition from variable-strength weak measurements}},\ }\href
  {https://doi.org/10.1103/PhysRevB.100.064204} {\bibfield  {journal} {\bibinfo
   {journal} {Phys. Rev. B}\ }\textbf {\bibinfo {volume} {100}},\ \bibinfo
  {pages} {064204} (\bibinfo {year} {2019})}\BibitemShut {NoStop}%
\bibitem [{\citenamefont {Szyniszewski}\ \emph {et~al.}(2020)\citenamefont
  {Szyniszewski}, \citenamefont {Romito},\ and\ \citenamefont
  {Schomerus}}]{Szyniszewski2020UniversalityEntanglement}%
  \BibitemOpen
  \bibfield  {author} {\bibinfo {author} {\bibfnamefont {M.}~\bibnamefont
  {Szyniszewski}}, \bibinfo {author} {\bibfnamefont {A.}~\bibnamefont
  {Romito}},\ and\ \bibinfo {author} {\bibfnamefont {H.}~\bibnamefont
  {Schomerus}},\ }\bibfield  {title} {\bibinfo {title} {Universality of
  entanglement transitions from stroboscopic to continuous measurements},\
  }\href {https://doi.org/10.1103/PhysRevLett.125.210602} {\bibfield  {journal}
  {\bibinfo  {journal} {Phys. Rev. Lett.}\ }\textbf {\bibinfo {volume} {125}},\
  \bibinfo {pages} {210602} (\bibinfo {year} {2020})}\BibitemShut {NoStop}%
\bibitem [{\citenamefont {Haake}\ \emph {et~al.}(2018)\citenamefont {Haake},
  \citenamefont {Gnutzmann},\ and\ \citenamefont {Ku\'{s}}}]{Haake2018QSoC}%
  \BibitemOpen
  \bibfield  {author} {\bibinfo {author} {\bibfnamefont {F.}~\bibnamefont
  {Haake}}, \bibinfo {author} {\bibfnamefont {S.}~\bibnamefont {Gnutzmann}},\
  and\ \bibinfo {author} {\bibfnamefont {M.}~\bibnamefont {Ku\'{s}}},\
  }\href@noop {} {\emph {\bibinfo {title} {Quantum Signatures of Chaos}}}\
  (\bibinfo  {publisher} {Springer, Berlin},\ \bibinfo {year}
  {2018})\BibitemShut {NoStop}%
\bibitem [{\citenamefont {Haake}\ \emph {et~al.}(1987)\citenamefont {Haake},
  \citenamefont {Ku{\'s}},\ and\ \citenamefont {Scharf}}]{haake1987classical}%
  \BibitemOpen
  \bibfield  {author} {\bibinfo {author} {\bibfnamefont {F.}~\bibnamefont
  {Haake}}, \bibinfo {author} {\bibfnamefont {M.}~\bibnamefont {Ku{\'s}}},\
  and\ \bibinfo {author} {\bibfnamefont {R.}~\bibnamefont {Scharf}},\
  }\bibfield  {title} {\bibinfo {title} {Classical and quantum chaos for a
  kicked top},\ }\href@noop {} {\bibfield  {journal} {\bibinfo  {journal} {Z.
  Phys. B}\ }\textbf {\bibinfo {volume} {65}},\ \bibinfo {pages} {381}
  (\bibinfo {year} {1987})}\BibitemShut {NoStop}%
\bibitem [{\citenamefont {Dyson}(1962)}]{Dyson1962Brownian}%
  \BibitemOpen
  \bibfield  {author} {\bibinfo {author} {\bibfnamefont {F.~J.}\ \bibnamefont
  {Dyson}},\ }\bibfield  {title} {\bibinfo {title} {A {Brownian}‐motion model
  for the eigenvalues of a random matrix},\ }\href
  {https://doi.org/10.1063/1.1703862} {\bibfield  {journal} {\bibinfo
  {journal} {J. Math. Phys.}\ }\textbf {\bibinfo {volume} {3}},\ \bibinfo
  {pages} {1191} (\bibinfo {year} {1962})}\BibitemShut {NoStop}%
\end{thebibliography}

%

\end{document}